\documentclass[a4paper,10pt,fleqn]{article}
\pdfoutput=1
\usepackage{jheppub}
\usepackage[T1]{fontenc}
\usepackage{setspace}
\usepackage{enumitem}
\usepackage{amsmath}
\usepackage{amsthm}
\usepackage{bm}
\usepackage{mathtools}
\usepackage{caption}
\usepackage{subcaption}
\usepackage{tikz}
\usepgflibrary{arrows}
\usetikzlibrary{calc}

\definecolor{jlab_red}{RGB}{192,39,45}
\definecolor{jlab_orange}{RGB}{249,102,0}
\definecolor{jlab_blue}{RGB}{47,122,121}
\definecolor{jlab_green}{RGB}{65,125,10}

\hypersetup{%
pdftitle = {Scattering from Euclidean},
pdfsubject = {QCD},
pdfauthor = {Bruno and Hansen},
colorlinks = {true},
filecolor = {black},
linkcolor = {jlab_blue},
menucolor = {black},
citecolor = {jlab_green},
urlcolor = {jlab_green},
}{}

\newcommand{\Sec}[0]{section}
\newcommand{\Secs}[0]{sections}
\newcommand{\Eq}[0]{eq.}
\newcommand{\Eqs}[0]{eqs.}
\newcommand{\Fig}[0]{figure}
\newcommand{\Reference}[0]{ref.}
\newcommand{\References}[0]{refs.}
\newcommand{\App}[0]{appendix}

\renewcommand{\vec}[0]{\boldsymbol}

\newcommand{\GenTwoThree}[0]{
Rummukainen:1995vs,
He:2005ey,
Kim:2005gf,
Lage:2009zv,
Bernard:2010fp,
Fu:2011xz,
Hansen:2012tf,
Briceno:2012yi,
Guo:2012hv,
Briceno:2012rv,
Polejaeva:2012ut,
Briceno:2014oea,
Hansen:2014eka,
Hansen:2015zga,
Briceno:2017tce,
Hammer:2017uqm,
Hammer:2017kms,
Mai:2017bge,
Briceno:2018aml,
Briceno:2018mlh,
Jackura:2019bmu,
Blanton:2019igq,
Briceno:2019muc,
Hansen:2019nir,
Romero-Lopez:2019qrt,
Blanton:2020gha,
Blanton:2020jnm,
Hansen:2020zhy,
Blanton:2020gmf}

\title{Variations on the Maiani-{\!}Testa approach and the inverse problem}
\author[a]{M.~Bruno}
\author[b]{and M.~T.~Hansen}
\affiliation[a]{Theoretical Physics Department, CERN, 1211 Geneva 23, Switzerland}
\affiliation[b]{Higgs Centre for Theoretical Physics, School of Physics and Astronomy, The University of Edinburgh, Edinburgh EH9 3FD, UK}
\emailAdd{mattia.bruno@cern.ch}
\emailAdd{maxwell.hansen@ed.ac.uk}

\preprint{CERN-TH-2020-412}

\abstract{We discuss a method to construct hadronic scattering and decay amplitudes from Euclidean correlators, by combining the approach of a regulated inverse Laplace transform with the work of Maiani and Testa \cite{Maiani:1990ca}. Revisiting the original result of \Reference~\cite{Maiani:1990ca}, we observe that the key observation, i.e.~that only threshold scattering information can be extracted at large separations, can be understood by interpreting the correlator as a spectral function, $\rho(\omega)$, convoluted with the Euclidean kernel, $e^{- \omega t}$, which is sharply peaked at threshold. We therefore consider a modification in which a smooth step function, equal to one above a target energy, is inserted in the spectral decomposition. This can be achieved either through Backus-Gilbert-like methods or more directly using the variational approach. The result is a shifted resolution function, such that the large $t$ limit projects onto scattering or decay amplitudes above threshold. The utility of this method is highlighted through large $t$ expansions of both three- and four-point functions that include leading terms proportional to the real and imaginary parts (separately) of the target observable. This work also presents new results relevant for the un-modified correlator at threshold, including expressions for extracting the $N \pi$ scattering length from four-point functions and a new strategy to organize the large $t$ expansion that exhibits better convergence than the expansion in powers of $1/t$.}

\begin{document}
\maketitle
\clearpage
\abovedisplayskip 11pt
\belowdisplayskip 11pt


\section{Introduction}

Over the last decades, numerical lattice QCD has become a high-precision tool for predicting several non-perturbative strong-force observables, including hadronic masses, decay constants and form factors. Looking beyond these quantities, each defined in terms of single-hadron states, lattice QCD has also shown outstanding progress in calculating multi-hadron observables including $2 \to 2$ scattering and $1 \to 2$ decay amplitudes. The successful determination of such amplitudes is especially impressive due to the presence of the Euclidean metric, required in order to apply Monte Carlo methods in estimating the lattice QCD path integral. While the analytic continuation to Minkowski correlators is formally guaranteed by the Osterwalder-Schrader theorem \cite{Osterwalder:1973dx}, in practice the limited knowledge of correlation functions on a finite set of points, together with the presence of statistical uncertainties, implies that direct extraction is a numerically ill-posed inverse problem, see e.g.~\Reference~\cite{Bertero}.

Offering another perspective on this challenge, Maiani and Testa \cite{Maiani:1990ca} showed that (for energies above production threshold) asymptotically separating individual pion fields in Euclidean time leads to correlators dominated by off-shell contributions. As we review in more detail in section~\ref{sec:3pt}, this means that matrix elements of the form $\langle \pi \vert \hat \pi(0) \vert \pi \pi \rangle$ contribute, where $\langle \pi \vert$ represents a single pion state and $\vert \pi \pi \rangle$ a two-pion asymptotic in- or out-state. The infamous off-shellness refers to the fact that the four momentum associated with the operator $\hat \pi$ does not satisfy $q^2 = m_\pi^2$, where $m_\pi$ is the physical particle mass. As a result the matrix element depends on the details of the operator used and gives no useful information.

This limitation was circumvented by L\"uscher \cite{Luscher:1985dn,Luscher:1986pf}, who showed that the values of the low-lying two-pion energies in a finite periodic spatial volume are determined by the on-shell $2\to2$ scattering amplitude (up to corrections falling faster than any power of $1/L$ where $L$ denotes the box length). Thus, the scattering amplitude can be extracted indirectly, from discrete spectra determined in lattice calculations. In more recent years, due especially to efficient methods for evaluating quark-field Wick contractions \cite{Peardon:2009gh} and an improved understanding of the importance of a large operator basis \cite{Blossier:2009kd}, this method has proven to be extremely successful in the determination of elastic two-hadron scattering amplitudes.\footnote{See \References~\cite{Briceno:2017max,Padmanath:2019wid,Bulava:2019hpz,Edwards:2020rbo} for recent reviews.} Formal extensions of the relations between amplitudes and energies \cite{\GenTwoThree} have allowed the same basic approach to be applied in calculations involving particles with spin, coupled-channel two-particle systems, and most recently to the scattering of three-pion states.

In parallel, Lellouch and L\"uscher \cite{Lellouch:2000pv} extended this approach, to give a method for extracting the $K \to (\pi\pi)_{I=0,2}$ decay amplitude from a combination of finite-volume energies and matrix elements. This approach has since been applied by the RBC-UKQCD collaboration \cite{Abbott:2020hxn,Bai:2015nea,Blum:2015ywa} to reach a first-principles understanding of the $\Delta I = 1/2$ rule and a determination of the CP violating parameter $\epsilon'/\epsilon$. Further generalizations to generic $0 \overset{\mathcal J}{\to} 2$, $1 \overset{\mathcal J}{\to} 2$ and $2 \overset{\mathcal J}{\to} 2$ amplitudes have been derived since \cite{Lin:2001ek,Detmold:2004qn,Kim:2005gf,Christ:2005gi,Meyer:2011um,Hansen:2012tf,Briceno:2012yi,Bernard:2012bi,Agadjanov:2014kha,Briceno:2014uqa,Feng:2014gba,Briceno:2015csa,Briceno:2015tza,Baroni:2018iau,Briceno:2019nns,Briceno:2020xxs,Feng:2020nqj} with the symbol above the arrow indicating that the transition is mediated by a local operator, e.g.~the conserved vector current.

The main limitation of the finite-volume formalism, besides the technical challenge of measuring several excited-state energy levels and matrix elements, is the proliferation of multi-particle channels as the scattering energy increases. Here it is important to note that any finite-volume method must formally treat all open multi-hadron channels (or argue that they are irrelevant) in order to reach a prediction about any individual scattering, decay, or transition amplitude. This is because finite-volume energies and matrix elements generically depend on a mixture of all physically allowed scattering processes.

For these reasons, recent work has revisited prospects for the direct analytic continuation of numerical correlation functions in the context of inverting the Laplace transform
\begin{equation}
\label{eq:Gtau}
G(t,L) = \int_0^\infty d \omega \, e^{- \omega t} \, \rho(\omega,L) \qquad \qquad (t>0) \,,
\end{equation}
where $G(t,L)$ represents a general Euclidean correlator and the right-hand side defines the spectral function, $ \rho(\omega,L)$. A regulated inverse of this simple relation could potentially unlock an alternative method in extracting inclusive quantities, such as heavy-particle lifetimes or the hadronic tensor \cite{Hansen:2017mnd}, as well as scattering amplitudes \cite{Bulava:2019kbi}, directly from spectral functions convoluted with a known resolution function
\begin{equation}
\label{eq:rhobar}
\widehat{\rho} (\overline \omega, \Delta, L) \equiv \int_0^\infty d \omega \,
\widehat \delta_{\Delta}(\omega, \overline \omega) \, \rho(\omega, L) \,,
\end{equation}
where $\widehat \delta_{\Delta}(\omega, \overline \omega)$ is peaked at $\omega = \overline \omega$ with characteristic width $\Delta$. Encouraging progress has recently been made in developing improved strategies to regulate the inverse problem and systematically target a resolution function~\cite{Hansen:2019idp,Bailas:2020qmv} to extract \Eq~\eqref{eq:rhobar} from \Eq~\eqref{eq:Gtau}. To reach a physical prediction, these methods formally require that the infinite-volume limit ($L \to \infty$) is taken before the resolution width is sent to zero ($\Delta \to 0$). Such ideas might prove useful also in the context of QED corrections to semi-leptonic decays, and similar processes,\footnote{%
Other notable examples are long distance effects in $\epsilon_K$ and time-like Compton amplitudes. See also \References~\cite{Christ:2015pwa} and \cite{Briceno:2019opb} for finite-volume methods targeting these observables.}
where intermediate on-shell states prevent the analytic continuation to Minkowski signature and approximate numerical solutions to the inverse problem could play a significant role.

In this manuscript, we present a strategy for combining the ideas summarized by \Eqs~\eqref{eq:Gtau} and \eqref{eq:rhobar} with the work of Maiani and Testa \cite{Maiani:1990ca}, in order to extend the reach of the latter to energies above scattering threshold, without suffering the dominance of off-shell terms. We present the idea both in the context of three- and four-point functions and, as a side benefit, we also reach new results for extracting threshold information from standard correlators.

Although we focus in this work on three- and four-point functions, the basic idea can already be expressed with a two point function of a scalar current, $J(t, \vec x)$,
\begin{equation}
G(t) = \int d^3 \vec x \, \langle J(t, \vec x) J(0) \rangle \,,
\end{equation}
where we assume the $L \to \infty$ limit has been taken. We then define the modified correlator
\begin{equation}
G^{\Theta}(t \vert s) = \int d^3 \vec x \, \langle J(t, \vec x) \Theta(\hat H - \sqrt{s}, \Delta) J(0) \rangle \,,
\end{equation}
where $\Theta(z, \Delta)$ is a smoothened Heaviside function, interpolating from zero for $z<0$ to one for $z>0$. A specific definition of $\Theta(z, \Delta)$ is given in \Eq~\eqref{eq:ThetaDef} below, but any function can be used provided it is smooth and becomes the usual Heaviside step function for $\Delta \to 0$. Defining the spectral function as
\begin{equation}
\rho(\omega) = \int d^3 \vec x \, \langle J(0, \vec x) \delta(\hat H - \omega) J(0) \rangle \,,
\end{equation}
note that the following relations hold:
\begin{equation}
G(t) = \int_0^\infty d \omega \, e^{- \omega t} \, \rho(\omega) \,, \qquad \qquad G^{\Theta}(t \vert s) = \int_0^\infty d \omega \, \big [ \Theta(\omega - \sqrt{s}, \Delta) e^{- \omega t} \big ] \, \rho(\omega) \,.
\end{equation}
These two simple results form the basis of this work. In the expression for $G(t)$, the kernel $e^{- \omega t}$ becomes sharply peaked at threshold for large $t$, leading to the threshold dominance famously identified by Maiani and Testa \cite{Maiani:1990ca}. For $G^{\Theta}(t \vert s)$, by contrast, the peak forms at $\omega = \sqrt{s}$ and one can extract time-like observables away from threshold.\footnote{We also point the reader to \References~\cite{Gambino:2020crt,Fukaya:2020wpp}, in which the authors also use a spectral representation involving a smooth theta function, but in a different context. The details of their intriguing proposal differ from those presented here, most importantly because, in \Reference~\cite{Gambino:2020crt,Fukaya:2020wpp}, the range of sampled energies is given a physical interpretation as a phase-space integral defining a scattering cross section. In the present work we consider amplitudes at fixed energies (rather than total rates) so that no phase-space integral arises.} See also \Fig~\ref{MTvsINV} below.

To give an impression of the relations derived in the following, we close this introduction by highlighting two of our key results.

First, returning to the scalar current, $J(x)$, we introduce the form factor
\begin{equation}
f(s) = \langle s, \pi \pi, \text{out} \vert J(0) \vert 0 \rangle \,,
\end{equation}
where $\vert 0 \rangle$ is the vacuum and $ \langle s, \pi \pi, \text{out} \vert$ is a two-particle out state with squared center-of-momentum energy, $s$, projected to zero angular momentum by the current. We then demonstrate in \Sec~\ref{sec:3pt} that the following holds
\begin{multline}
\label{eq:3ptThetaIntro}
\mathcal N \, \langle \pi , \vec q \vert \widetilde \pi_{- \vec q}(t) \, \Theta(\hat H - \sqrt{s}, \Delta) \, J(0) \vert 0 \rangle \\ = e^{-\sqrt{m_\pi^2 + \vec q^2} t} \bigg[ \Theta(0,\Delta) \, \text{Re} [ f ( s ) ] -2 \mathcal J^{(0)}(t, s, \Delta) \, \text{Im} [ f ( s ) ] + \cdots \bigg]_{s = 4(\vec q^2 + m_\pi^2)} \,.
\end{multline}
Here the left-hand side is a product of a simple normalization factor, $\mathcal N$, with an infinite-volume Euclidean matrix element, modified by the smooth Heaviside-function $\Theta (z, \Delta)$. The matrix element is built from a single-pion state and a generic operator $ \widetilde \pi_{- \vec q}(0)$ such that $\langle 0 \vert \widetilde \pi_{- \vec q}(0)$ has the quantum numbers of $\langle \pi, - \vec q \vert$.\footnote{We comment here that, in \Reference~\cite{Lin:2001ek}, the authors also consider methods that do not make use of the Lellouch-L\"uscher type conversion factor. These differ from the present work, for example in that the observable is still extracted from a single finite-volume matrix element.}

The right-hand side of \Eq~\eqref{eq:3ptThetaIntro} demonstrates the utility of this quantity. Specifically, as we prove in \Sec~\ref{sec:3pt}, it is equal to a known linear combination (with one time-independent coefficient $\Theta(0,\Delta)$ and a second time-dependent function $\mathcal J^{(0)}$), of the real and imaginary parts of the target observable, $f(s)$, up to terms that are suppressed for well-chosen values of $t$ and $\Delta$. The suppression is quantified through an asymptotic series of known geometric functions that can be understood as a generalization of the large $t$ expansion of Maiani and Testa \cite{Maiani:1990ca}. As with the inverse techniques of \References~\cite{Hansen:2017mnd,Bulava:2019kbi}, this result holds for all $s$ and may be competitive with standard methods for $s>(4 m_\pi)^2$ ($m_\pi$ is the pion mass), where it is challenging to disentangle the multiple open channels from finite-volume information.

A simple cross check of \Eq~\eqref{eq:3ptThetaIntro} is given by setting $s = 4 m_\pi^2$ and carefully taking the $\Delta \to 0$ limit, as detailed in \Sec~\ref{sec:Threshold3pt}. Then one recovers the original result of \Reference~\cite{Maiani:1990ca}
\begin{equation}
\mathcal N \, \langle \pi , \vec 0 \vert \, \widetilde \pi_{ \vec 0}(t) \, J(0) \, \vert 0 \rangle = e^{- m_\pi t} f ( 4 m_\pi^2 ) \big [ 1 + \mathcal O(t^{-1/2}) \big ] \,.
\end{equation}
In \Sec~\ref{sec:Threshold3pt} we also discuss the finite $t$ corrections to this and revisit the expansion performed by Maiani and Testa. We express the result in an alternative basis that exhibits faster convergence and give expressions for the next-to-leading order term (depending on the $\pi \pi$ scattering length) and the next-to-next-to-leading order term (depending on the derivative of the scattering amplitude with respect to the virtuality of an off-shell leg).

Second, we consider the same derivation with four-point functions, studying both the standard correlation function and the modification with $\Theta$. In direct analogy to \Eq~\eqref{eq:3ptThetaIntro} above, we deduce that one can extract the $2 \to 2$ scattering amplitude at all energies from the $\Theta$-correlator. Also of great interest, however, is the standard correlation function at two-particle threshold, which is more straight-forward to implement in the short term. For example in the case of $N \pi$ scattering, we derive the following result in \Sec~\ref{sec:4pt}
\begin{multline}
\mathcal N^2 \, C_{a'b'} C_{ab} \,
\langle \pi^{a'} , \vec 0 \vert \, \widetilde N^{b'}_{ \vec 0}(t) \, \widetilde N^b_{\vec 0}(0) \, \vert \pi^{a}, \vec 0 \rangle_{\sf c} \\[5pt]
= e^{-m_N t} \bigg[ 8 \pi (m_N + m_\pi) \, a_{N\pi} \, t
- 16 \, a_{N \pi}^2 \sqrt{2 \pi (m_N + m_\pi) m_N m_\pi t}
+ \mathcal O \big (t^0\big) \bigg] \,.
\label{eq:NpiThreshIntro}
\end{multline}
Here the left-hand side is a Euclidean matrix element that can be extracted from a four-point function of pion and nucleon fields, all projected to zero spatial momentum. The indices $a',a$ encode isospin and $b',b$ simultaneously indicate isospin and spin such that the coefficients $C_{ab}$ can be chosen to project onto any definite $N \pi$ quantum numbers (with an implicit sum over indices understood). The left-hand side also includes a simple normalization factor $\mathcal N^2$ and the subscript ${\sf c}$ indicates that the disconnected $N \to N$ and $\pi \to \pi$ contribution is subtracted.

In analogy to the Maiani-Testa correlation function at threshold, this four-point function can be used to directly extract the $N \pi$ scattering length, $a_{N \pi}$, in the channel specified by $ C_{ab}$. The advantage compared to the three-point function is that one recovers a function only of $a_{N\pi}$ as opposed to a combination of the threshold form-factor [$f(4 m_\pi^2)$] and the scattering length. Indeed the right-hand side contains two terms, scaling as $t a_{N \pi}$ and $\sqrt{t} a_{N \pi}^2$ so that a strong constraint on the scattering length may be achieved from the combined fit.

The remainder of this manuscript is organized as follows: In the next section we revisit the derivation of ref.~\cite{Maiani:1990ca} and show how the result is modified for the $\Theta$-correlator. We also present a new form of the large $t$ expansion, in terms of a series of known integrals. We then repeat the analysis for four-point functions in \Sec~\ref{sec:4pt}, where we also present the derivation of eq.~\eqref{eq:NpiThreshIntro}. In \Sec~\ref{sec:imp}, we discuss implementation strategies to extract the $\Theta$-correlator, including the role of finite-volume effects. We also include a number of appendices, deriving various technical results used in the main text.


\section{Time-like form factors\label{sec:3pt}}

Following Maiani and Testa we begin by defining the Euclidean correlator
\begin{align}
G_{\vec q_1 \vec q_2}(t_1,t_2) & \equiv \big \langle \widetilde \pi_{\vec q_1}(t_1) \widetilde \pi_{ \vec q_2}(t_2) J(0) \big \rangle \,,
\end{align}
where $\widetilde \pi_{\vec q}(t) = \int d^3 \vec x \, \pi(t,\vec x) e^{- i \vec q \cdot \vec x}$ is a generic single-pion interpolating field and $J(0)$ is a local scalar current. Then, taking $t_1$ large and positive, we write
\begin{align}
G_{\vec q_1 \vec q_2}(t_1, t_2) & = Z_{\pi} \, \frac{e^{- \omega_{\vec q_1} t_1 }}{ 2 \omega_{\vec q_1}} \frac{e^{- \omega_{\vec q_2} t_2 }}{2 \omega_{\vec q_2}} c_{\vec q_1 \vec q_2}(t_2) + \mathcal O \Big (e^{- \sqrt{9 m_\pi^2 + \vec q_1^2} t_1} \Big ) \,, \\
c_{\vec q_1 \vec q_2}(t_2) & \equiv \frac{ 2 \omega_{\vec q_2} e^{\omega_{\vec q_2} t_2 } \, \langle \pi , \vec q_1 \vert \widetilde \pi_{ \vec q_2}(t_2) J(0) \vert 0 \rangle }{\sqrt{Z_\pi}} \,,
\end{align}
where $\omega_{\vec q} = \sqrt{\vec q^2 + m_\pi^2}$ and $Z_\pi = \langle \pi, \vec p \vert \pi(0) \vert 0 \rangle^2$. Here the single particle state has the usual relativistic normalization, $\langle \pi, \vec q \vert \pi, \vec q' \rangle = 2 \omega_{\vec q} (2 \pi)^3 \delta^3(\vec q - \vec q')$.

As was shown in \Reference~\cite{Maiani:1990ca}, for $\vec q_1 \neq \vec q_2$ [which, for vanishing total momentum ($\vec q_1 + \vec q_2= \vec 0$), translates to $\vec q_1 \neq \vec 0$], the quantity $c_{\vec q_1 \vec q_2}(t_2)$ contains growing exponentials as $t_2 \to \infty$, with coefficients depending on scattering amplitudes for which one of the external legs is off the mass shell. To better understand the limitations associated with this, we define the modified quantity
\begin{align}
\label{eq:cThetaDef}
c^{\Theta}_{\vec q_1 \vec q_2}(t_2 \vert \omega_0) & \equiv \frac{ 2 \omega_{\vec q_2} e^{\omega_{\vec q_2} t_2 } \, \langle \pi , \vec q_1 \vert \widetilde \pi_{ \vec q_2}(0) \, \Theta(\hat M - 2 \omega_0, \Delta) \, e^{- (\hat H - \omega_{\vec q_1}) t_2} \, J(0) \vert 0 \rangle }{\sqrt{Z_\pi}} \,,
\end{align}
where $\Theta(\omega, \Delta)$ is a smooth Heaviside function, e.g.~
\begin{equation}
\Theta(\omega,\Delta) = \frac{\tanh(2 \omega/\Delta)+1}{2} \,,
\label{eq:ThetaDef}
\end{equation}
and $\hat M = \sqrt{ \hat H^2 - \hat P^2}$ is an operator giving the center-of-momentum-frame energy. In \Sec~\ref{sec:imp}, we describe strategies for accessing this object in a numerical lattice calculation, including the role of the finite-volume boundary conditions. In this section we take $c^{\Theta}$ as given and show that it can be used to access the real and imaginary parts of
\begin{equation}
f\big ( s(\vec q) \big ) \equiv \langle \pi \pi, \text{out}, \vec q_1 \vec q_2 \vert J(0) \vert 0 \rangle \,,
\end{equation}
where we have introduced
\begin{equation}
s(\vec q) \equiv ( \omega_{\vec q_1} + \omega_{\vec q_2} )^2 - (\vec q_1 + \vec q_2)^2 \,.
\end{equation}
In the same spirit as the inverse methods described in \References~\cite{Hansen:2017mnd,Bulava:2019kbi}, this approach is valid for all values of $s(\vec q)$, provided the correlator can be estimated with an appropriate range of $\Delta$ and $\omega_0$.

\begin{figure}
\begin{center}
\includegraphics[width=\textwidth]{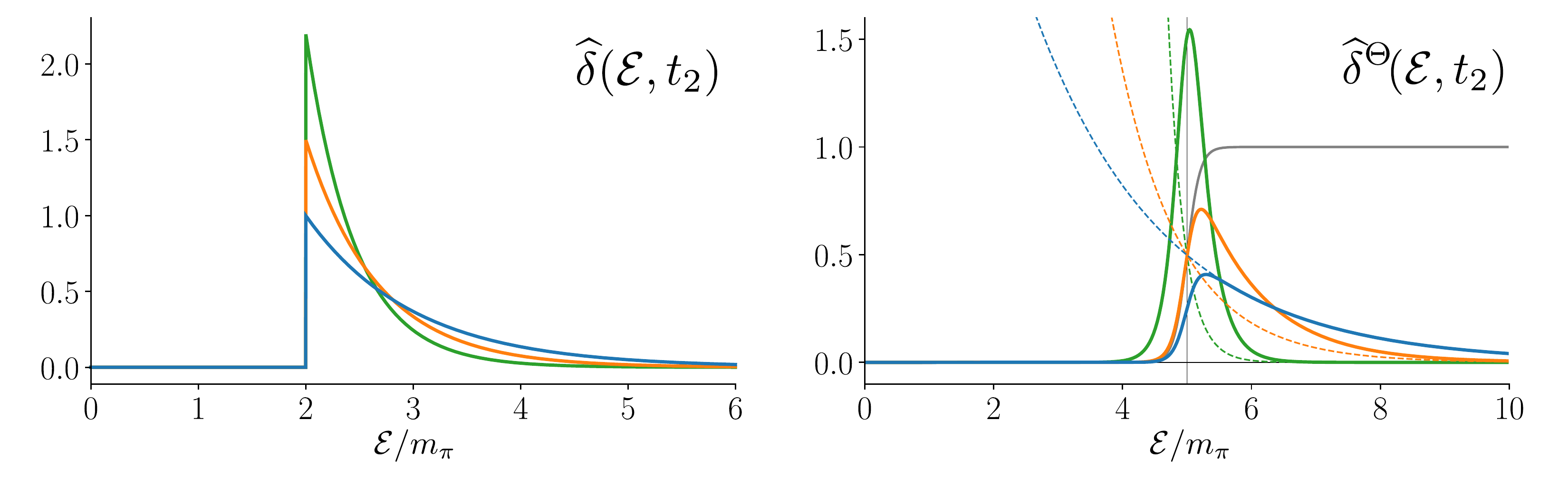}
\caption{Sketch of the normalized effective resolution functions, $\widehat \delta(\mathcal E,t_2)$ and $\widehat \delta^\Theta(\mathcal E,t_2)$, defined in eqs.~\eqref{eq:deltahat} and \eqref{eq:deltahatTheta} respectively. The corresponding correlators can be written as convolution integrals of these resolution functions with the same spectral function, $\rho_{\vec q}(\mathcal E)$, which contains the time-like information we are after. Thus, the functional forms plotted here give an indication of the energies that predominantly contribute to the correlation functions. The left panel illustrates that $c_{\vec q,-\vec q}(t_2)$ is dominated by $\mathcal E \approx 2 m_\pi$ and that large $t_2$ sharpens the near-threshold resolution. By contrast, the right panel illustrates the modification involving $\Theta(\mathcal E - 2\omega_{\vec q}, \Delta)$ (also plotted here in gray) for $2\omega_{\vec q}=5 m_\pi$. The three curves correspond to the same $\Delta$ value, and the sharpening around $\mathcal E = 5 m_\pi$ is achieved by increasing $t_2$, with the dashed curves showing the corresponding factors of $e^{- t_2 ( \mathcal E - 2 \omega_{\vec q})}$. \label{MTvsINV}}
\end{center}
\end{figure}

Before deriving our main result, we pause here to discuss the intuition motivating eq.~\eqref{eq:cThetaDef} and its relation to the results of Maiani and Testa. For this discussion we set $\vec q_1 = - \vec q_2 = \vec q$ and $\omega_0=\omega_{\vec q}$. The key point is that $c_{\vec q, -\vec q}(t_2)$ and $c^{\Theta}_{\vec q, -\vec q}(t_2 \vert \omega_{\vec q})$ can be written as the convolution of the same spectral function, $\rho$, with two different resolution functions, $\widehat \delta$ and $\widehat \delta^{\Theta}$. In particular
\begin{align}
c_{\vec q, -\vec q}(t_2) & = \int_0^\infty d \mathcal E \, \rho_{\vec q}(\mathcal E) \, \widehat \delta(\mathcal E, t_2) \,, \\
c^{\Theta}_{\vec q, -\vec q}(t_2 \vert \omega_{\vec q}) & = \int_0^\infty d \mathcal E \, \rho_{\vec q}(\mathcal E) \, \widehat \delta^{\Theta}(\mathcal E, t_2) \,,
\end{align}
where
\begin{align}
\rho_{\vec q}(\mathcal E) & = \frac{ 2 \omega_{\vec q} \, \langle \pi , \vec q \vert \widetilde \pi_{- \vec q}(0) \, \delta(\hat H - \mathcal E ) \, J(0) \vert 0 \rangle }{\sqrt{Z_\pi}} \,, \\[3pt]
\label{eq:deltahat}
\widehat \delta(\mathcal E, t_2) & = \theta(\mathcal E - 2 m_\pi) \, e^{- (\mathcal E - 2\omega_{\vec q}) t_2} \,, \\[6pt]
\label{eq:deltahatTheta}
\widehat \delta^{\Theta}(\mathcal E, t_2)& = \Theta(\mathcal E - 2 \omega_{\vec q}, \Delta) \, e^{- (\mathcal E -2\omega_{\vec q})t_2} \,.
\end{align}
Here we have included a zero-width Heaviside function (denoted $\theta(x)$) in the definition of $\widehat \delta(\mathcal E, t_2)$ to further emphasize the similarities. This is allowed as the spectral function, $\rho(\mathcal E)$, has zero support for $\mathcal E < 2 m_\pi$.\footnote{We assume throughout that the system has no bound state.}

In \Fig~\ref{MTvsINV} we show the functional forms of $\widehat \delta(\mathcal E, t_2) $ and $\widehat \delta^{\Theta}(\mathcal E, t_2)$, normalized to unit area. Note that $\widehat \delta(\mathcal E, t_2)$ is sharply peaked at threshold with a width given by $1/t_2$. For this reason, the large $t_2$ limit can only access time-like information at threshold, as was famously established in \Reference~\cite{Maiani:1990ca}. For $\widehat \delta^{\Theta}(\mathcal E, t_2)$, the peak is shifted and mimics the resolution functions discussed in refs.~\cite{Hansen:2017mnd,Bulava:2019kbi}, formally allowing one to extract scattering amplitudes at all energies. The key advantage, as compared to the earlier work, is that the effective resolution width can be reduced at fixed $\Delta$ by increasing $t_2$. This gives a powerful handle on the target observable and can be expressed as a large $t_2$ expansion, to which we now turn.

Returning to eq.~\eqref{eq:cThetaDef}, the next step is to insert a complete set of states between $\widetilde \pi_{\vec q_2}(0)$ and $J(0)$ to reach
\begin{equation}
\label{eq:CThetatoAB}
c^{\Theta}_{\vec q_1 \vec q_2}(t_2 \vert \omega_0) = \sum_k \int \! d \Phi_k \, e^{- ( E(\vec p) - \omega_{\vec q_1} - \omega_{\vec q_2}) t_2 } \, \Theta\big ( s(\vec p)^{1/2} - 2 \omega_0, \Delta \big ) A_k(\vec q_1, \vec q_2; \vec p ) B_k( \vec p ) \,,
\end{equation}
where
\begin{equation}
d \Phi_k = \frac{1}{S_k} \frac{d^3 \vec p_1}{(2 \pi)^3 2 \omega_{\vec p_1}} \cdots \frac{d^3 \vec p_{n_k}}{(2 \pi)^3 2 \omega_{\vec p_{n_k}}} (2 \pi)^3 \delta^3(\vec P - \vec q_1 - \vec q_2) \,,
\end{equation}
and $s(\vec p) = E(\vec p)^2 - \vec P^2$, with the total energy and momentum given by
\begin{equation}
\big ( E(\vec p), \vec P \big ) = \big ( \omega_{\vec p_1} + \cdots + \omega_{\vec p_{n_k}}, \, \vec p_1 + \cdots + \vec p_{n_k} \big ) \,.
\end{equation}
The complete set of states leads to the matrix elements
\begin{align}
A_k(\vec q_1, \vec q_2; \vec p) & \equiv 2 \omega_{\vec q_2} \frac{\langle \pi , \vec q_1 \vert \pi(0) \vert k, \text{out}, \vec p_1 \cdots \vec p_{n_k} \rangle}{\sqrt{Z_\pi}} \,, \\
B_k(\vec p) & \equiv \langle k, \text{out}, \vec p_1 \cdots \vec p_{n_k} \vert J(0) \vert 0 \rangle \,,
\end{align}
where $k$ is a channel index (referring e.g.~to $\pi \pi$, $K \overline K$, $\pi \pi \pi \pi$), $n_k$ is the number of particles in the channel and $S_k$ the number of permutations of identical particles. We define $k=1$ as the $\pi \pi$ channel, with internal quantum numbers matching $\pi(0) \vert \pi, \vec q \rangle$.

The final key step is to express $A_k(\vec q_1, \vec q_2; \vec p)$ in terms of an off-shell amplitude $\boldsymbol {\mathcal M}_k(\eta \vert \vec \mu(\vec q, \vec p))$. We use the bold notation here whenever the amplitude has at least one off-shell external leg. This object is defined in more detail in \App~\ref{app:ABproof} and for the present argument it suffices to note that $\boldsymbol {\mathcal M}_k(0 \vert \vec \mu(\vec q, \vec p)) = \mathcal M_k(\vec \mu(\vec q, \vec p))$ is the usual on-shell two-to-$n_k$ scattering amplitude as a function of Lorentz invariants organized in the vector $\vec \mu(\vec q, \vec p)$. For example in the case of a two-to-two amplitude
\begin{align}
\vec \mu (\vec q, \vec p) & = \big (s, t \big ) \,, \\
& = \Big ( (\omega_{\vec p_1} + \omega_{\vec p_2})^2 - (\vec p_1 + \vec p_2)^2 \,, \ (\omega_{\vec p_1} - \omega_{\vec q_1})^2 - (\vec p_1 - \vec q_1)^2 \Big ) \,.
\end{align}
The matrix element $A_k(\vec q_1, \vec q_2; \vec p)$ admits a simple decomposition in terms of the off-shell amplitude
\begin{equation}
A_k(\vec q_1, \vec q_2; \vec p)= \delta_{k,1} (2 \pi)^3 2 \omega_{\vec q_1} 2 \omega_{\vec q_2} \big [ \delta^3(\vec p_1 - \vec q_1) + \delta^3(\vec p_2 - \vec q_1) \big ] - 2 \omega_{\vec q_2} \frac{\boldsymbol {\mathcal M}_k\big (\eta(\vec q, \vec p) \vert \, \vec \mu(\vec q, \vec p) \big )^{\!*}}{ \eta( \vec q, \vec p) - i \epsilon} \,,
\label{eq:Afinal}
\end{equation}
where the distance of the $\pi(0)$ leg from the mass shell is given by
\begin{align}
\eta (\vec q, \vec p) & = ( E(\vec p)- \omega_{\vec q_1})^2 - (\vec P - \vec q_1)^2 - m_\pi^2 \,.
\label{eq:etazetadef}
\end{align}

Combining Eqs.~\eqref{eq:CThetatoAB} and \eqref{eq:Afinal} gives the generalization of Maiani and Testa's result to the case of non-zero total momenta and to the $\Theta$-function modification of the correlator. In \App~\ref{app:kinematics} we give explicit expressions in the case of general kinematics and discuss their utility. Here we focus on the role of the Heaviside function and set $\vec q_1 = - \vec q_2 = \vec q$ to reach
\begin{multline}
\label{eq:CThetatoABv2}
c^{\Theta}_{\vec q, - \vec q}(t_2 \vert \omega_0) = \Theta(2 \omega_q - 2 \omega_0, \Delta) \, f\big ( s( q^2) \big) \\
- 2 \omega_q \int_{2 m_\pi}^\infty \! \frac{d \mathcal E}{2 \pi} \, e^{- ( \mathcal E - 2 \omega_{q}) t_2 } \, \Theta( \mathcal E - 2 \omega_0, \Delta) \,
\frac{ \mathcal G(\mathcal E, \omega_q)}{ (\mathcal E - \omega_q)^2 - \omega_q^2 - i \epsilon} \,,
\end{multline}
where we have replaced $\vec q \to q$ in quantities that only depend on $\vec q^2 = q^2$ and have introduced
\begin{equation}
\label{eq:calGdef}
\mathcal G(\mathcal E, \omega_q) = \sum_k \int d \Phi_k \, 2 \pi \delta \big( \mathcal E - E(\vec p) \big ) \, \boldsymbol{\mathcal M}_k\big (\eta(\vec q, \vec p) \vert \, \vec \mu(\vec q, \vec p) \big )^{\!*} B_k(\vec p) \,.
\end{equation}

For example, in the case where only the $\pi \pi$ channel contributes, i.e.~for $\mathcal E^2 < (4 m_\pi)^2$, this becomes
\begin{align}
\mathcal G(\mathcal E, \omega_q) & = 2 \pi \frac12 \frac{4 \pi}{(2 \pi)^3} \int \! dp \, p^2 \, \frac{1}{4 \omega_{p}^2} \, \delta \big( \mathcal E - 2 \omega_{p} \big ) \, \boldsymbol{\mathcal M}_{\sf s}\big (\eta( q, p) \vert \, \mathcal E^2 \big )^{\!*} f(\mathcal E^2) \,, \\
& = \frac{ \sqrt{\mathcal E^2/4 - m_\pi^2} }{8 \pi \mathcal E} \, \boldsymbol{\mathcal M}_{\sf s}\big ((\mathcal E - \omega_q)^2 - \omega_q^2 \vert \, \mathcal E^2 \big )^{\!*} f(\mathcal E^2) \,,
\end{align}
where $\boldsymbol{\mathcal M}_{\sf s}$ is the scattering amplitude projected to zero orbital angular momentum
\begin{align}
\boldsymbol{\mathcal M}_{\sf s} \big(\eta(q,p) \vert \, \mathcal E^2 \big)= \frac{1}{4\pi} \int d\Omega \, \boldsymbol{\mathcal M}_1 \big( \eta(\vec q,\vec p) \vert \, \vec \mu(\vec q, \vec p) \big ) \,.
\end{align}
The on-shell limit is given by
\begin{align}
\mathcal G(\mathcal E, \mathcal E/2) & = \frac{ \sqrt{\mathcal E^2/4 - m_\pi^2} }{8 \pi \mathcal E} \frac{16 \pi \mathcal E}{ \sqrt{\mathcal E^2/4 - m_\pi^2} } \frac{e^{- 2 i \delta_{\sf s}(\mathcal E)} - 1}{- 2 i} e^{i \delta_{\sf s}(\mathcal E)} \vert f(\mathcal E^2) \vert \,, \\
& = 2 \sin \delta_{\sf s}(\mathcal E) \vert f(\mathcal E^2) \vert = 2 \, \text{Im} f(\mathcal E^2) \,.
\label{eq:Goptical}
\end{align}
Remarkably, the relation between the on-shell $\mathcal G$ function and $\text{Im} f(\mathcal E^2)$ in fact holds for all $\mathcal E^2$, as we prove in \App~\ref{app:optical}.

Returning to the general case, the final step is to perform the change of variables $x = ( \mathcal E - 2 \omega_q)t_2$ to reach
\begin{equation}
\label{eq:CThetaFinal}
c^{\Theta}_{\vec q, - \vec q}(t_2 \vert \omega_0) = \Theta(2 \omega_q - 2 \omega_0, \Delta) \, \text{Re} \Big [ f\big ( s( q^2) \big) \Big ]
- \int_{- (2 \omega_q - 2 m_\pi )t_2}^\infty \! dx \, \kappa(t_2, x) \, \mathcal G(x/t_2 + 2 \omega_q, \omega_q) \,,
\end{equation}
where
\begin{align}
\kappa(t_2, x) & = \frac{ \omega_q }{\pi} \, e^{-x } \,
\frac{ \Theta( x/t_2 + 2 \omega_q - 2 \omega_0, \Delta)}{x/t_2 + 2 \omega_q} \,
\mathcal P \frac{ 1}{ x } \,,
\label{eq:kappa}
\end{align}
and $\mathcal P$ indicates the principal-value pole prescription. To reach this expression, one uses the fact that $c^\Theta$ is real so that one can take the real part of the right-hand side for free. As $\mathcal G$ is also real this simply replaces: $ f ( s( q^2) ) \to \text{Re} f ( s( q^2) ) $ and $1/{(\eta - i \epsilon)} \to \mathcal P [1/\eta]$. Alternatively, one can demonstrate that the imaginary part of $1/{(\eta - i \epsilon)}$ explicitly cancels that of $f ( s( q^2) )$, as we review in \App~\ref{app:cancel3}, following \Reference~\cite{Maiani:1990ca}.

Equations \eqref{eq:calGdef}, \eqref{eq:CThetaFinal} and \eqref{eq:kappa} summarize the main new technical results of this section. In the following subsections we explore their implications for extracting scattering information.

\subsection{Threshold kinematics} \label{sec:Threshold3pt}

\begin{figure}
\begin{center}
\includegraphics[width=0.7\textwidth]{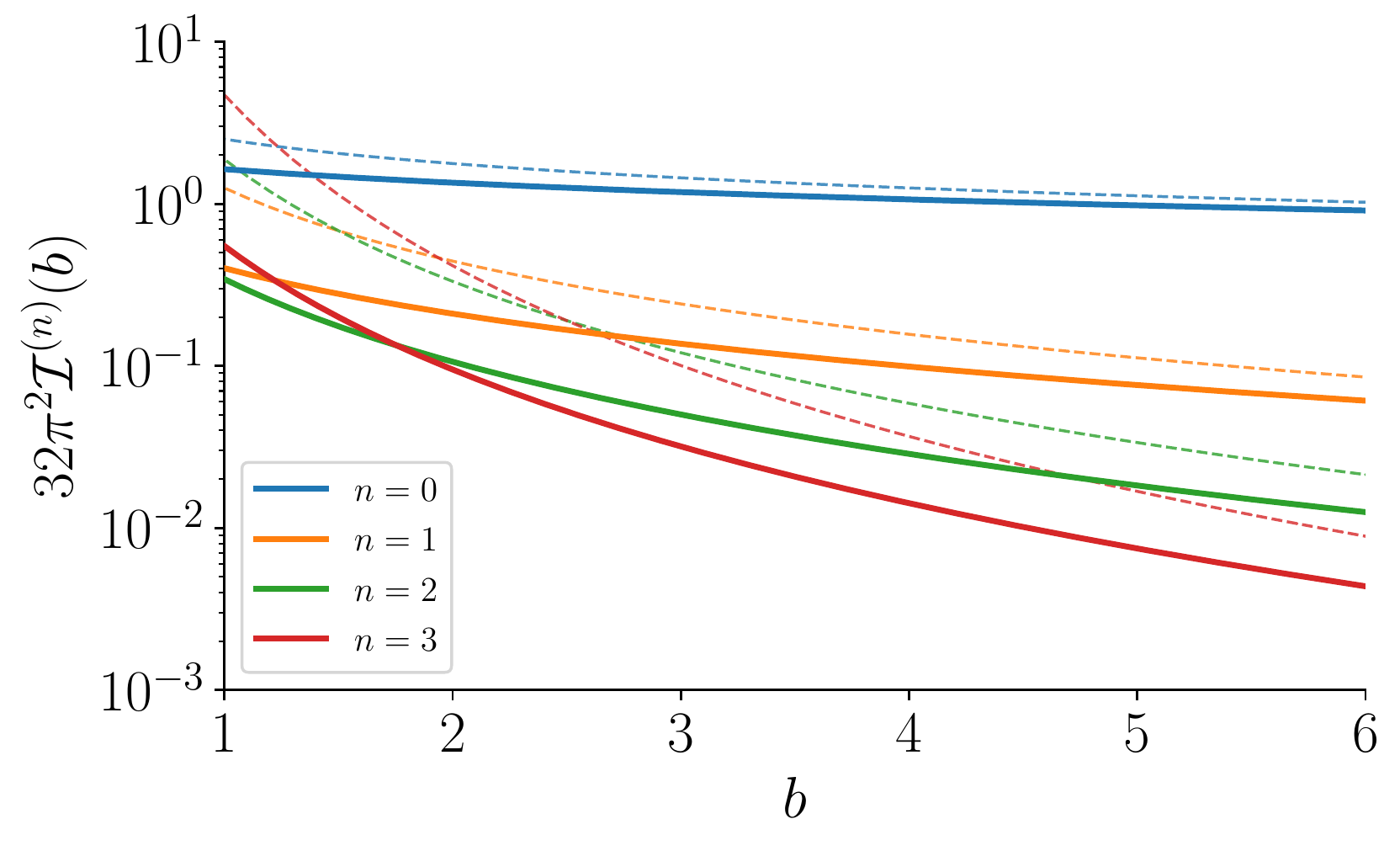}
\caption{Plot of the first few $\mathcal I^{(n)}(b)$ functions (solid lines), defined in \Eq~\eqref{eq:Idef} of the main text. The dashed lines show the large $b$ expansion for each function, scaling as $b^{-n-1/2}$ as shown in \Eq~\eqref{eq:Iexp}. \label{Iints}}
\end{center}
\end{figure}

As a first step we set $\omega_0 = \omega_q - \epsilon = m_\pi - \epsilon $ to reach the threshold case already considered in \Reference~\cite{Maiani:1990ca}. For these choices (together with $\Delta \to 0$) the $\Theta$-function has no effect and only the two-pion channel contributes in $\mathcal G$. We reach
\begin{equation}
\label{eq:CThetaThresh}
c^\Theta_{\,\vec 0, \vec 0}(t_2 \vert m_\pi) = f ( 4 m_\pi^2 )
\bigg[ 1 -
\sum_{n=0}^\infty g_n \, \mathcal I^{(n)}(2 m_\pi t_2) \bigg ] \,,
\end{equation}
or, to give an explicit expression in terms of the Euclidean correlator,
\begin{equation}
\big \langle \widetilde \pi_{\vec 0}(t_1) \widetilde \pi_{ \vec 0}(t_2) J(0) \big \rangle = Z_{\pi} \, \frac{e^{- m_\pi (t_1 +t_2) }}{ 4 m_\pi^2} f ( 4 m_\pi^2 ) \bigg[ 1 - \sum_{n=0}^\infty g_n \, \mathcal I^{(n)}(2 m_\pi t_2) \bigg ] \,,
\end{equation}
where $e^{- 3 m_\pi t_1}$ has been dropped. Here the $g_n$ coefficients arise in an expansion of the amplitudes
\begin{align}
\begin{split}
g_n & \equiv \frac{1}{n!} m_\pi^n \frac{d^n}{d\omega^n} \bigg [ \frac{ f ( 4 \omega^2 ) }{f(4 m_\pi^2)} \boldsymbol{\mathcal M}_{\sf s} \big ( (2 \omega - m_\pi)^2 - m_\pi^2 \vert \, 4 \omega^2 \big )^{\!*} \bigg ]_{\omega = m_\pi} \,,
\end{split}
\end{align}
and the functions $\mathcal I^{(n)}(b)$ are integrals of the kernel $\kappa$ times the phase space, $\sqrt{\mathcal E^2/4 - m_\pi^2}/(8 \pi \mathcal E)$, both in the limiting case of the threshold amplitude. They can be expressed as follows
\begin{align}
\label{eq:Idef}
\mathcal I^{(n)}(b)& \equiv \frac{1}{b^{n+1/2}} \int_0^\infty dx \, x^n \, \frac{ e^{- x }}{32 \pi^2 } \frac{ \sqrt{2 + x/b }}{ \sqrt{x} (1+x/b)^2 } \,,\\
& = \frac{\sqrt{2}}{32 \pi^2} \frac{1}{b^{n+1/2}} \Gamma(n+1/2) \Big [ 1 + \mathcal O \big ( 1/b \big ) \Big ] \,.
\label{eq:Iexp}
\end{align}
Expanding in powers of $1/b = 1/(2 m_\pi t_2)$ generates the original series presented in ref.~\cite{Maiani:1990ca}. However, as we illustrate in \Fig~\ref{Iints}, the leading terms converge slowly to the full integrals and the latter may prove to be a better basis in describing the correlation function.

Combining $g_0 = \mathcal M_{\sf s}(4 m_\pi^2) = 32 \pi m_\pi a_{\pi \pi}$ with $\Gamma(1/2) = \sqrt{\pi}$ gives the known result\footnote{Here we differ from eq.~(10) of \Reference~\cite{Maiani:1990ca} by a factor of 2 in the $1/\sqrt{t_2}$-dependent term. We are currently in correspondence with the authors of that work to clarify this difference. Note also that, following \Reference~\cite{Maiani:1990ca}, we have set our convention for the scattering-length such that $a_{\pi \pi} < 0$ for repulsive interactions at threshold.}
\begin{align}
\label{eq:CThreshExp}
c^\Theta_{\,\vec 0, \vec 0}(t_2 \vert m_\pi) & = f\big ( 4 m_\pi^2 \big) \bigg [ 1
- a_{\pi \pi} \sqrt{\frac{m_\pi}{ \pi t_2}} + \mathcal O \big (t_2^{-3/2} \big ) \bigg ] \,,
\end{align}
and thus
\begin{equation}
\big \langle \widetilde \pi_{\vec 0}(t_1) \widetilde \pi_{ \vec 0}(t_2) J(0) \big \rangle = Z_{\pi} \, \frac{e^{- m_\pi (t_1 +t_2) }}{ 4 m_\pi^2} f\big ( 4 m_\pi^2 \big) \bigg [ 1 - a_{\pi \pi} \sqrt{\frac{m_\pi}{ \pi t_2}} + \mathcal O \big (t_2^{-3/2} \big ) \bigg ] \,,
\end{equation}
where $a_{\pi\pi}$ is the two-particle scattering length. Our expressions allow one to easily reach generalizations of this, for example
\begin{equation}
\label{eq:CThreshExpAlt}
c^\Theta_{\,\vec 0, \vec 0}(t_2 \vert m_\pi) = f\big ( 4 m_\pi^2 \big) \bigg [ 1 - 32 \pi m_\pi a_{\pi \pi} \mathcal I^{(0)}(2 m_\pi t_2) - g_1 \mathcal I^{(1)}(2 m_\pi t_2) + \mathcal O \big (\mathcal I^{(2)} \big ) \bigg ] \,,
\end{equation}
with
\begin{equation}
\label{eq:g1Thresh}
g_1 = 256 \pi m_\pi^3 a_{\pi \pi} \frac{ \partial_s f ( s ) }{f(4 m_\pi^2)} + 8 m_\pi^2 \, \partial_s \mathcal M_{\sf s} ( s )^{*} + 4 m_\pi^2 \, \partial_\eta \boldsymbol{\mathcal M}_{\sf s} ( \eta \vert 4 m_\pi^2 )^{*} \bigg \vert_{s = 4 m_\pi^2\,, \ \eta = 0} \,.
\end{equation}
This gives a more explicit expression for the leading off-shellness contaminating this correlator. The $\eta$ derivative represents a small variation in the virtuality of one external leg away from the mass shell. The quantity is perfectly well defined but depends on the details of the operator $\pi(x)$ and has no physical meaning for pion scattering.

\subsection{General kinematics} \label{sec:General3pt}

Next we take $\omega_0 = \omega_q > m_\pi$ and write
\begin{equation}
\label{eq:CThetaGenEn}
c^\Theta_{\,\vec q, -\vec q}(t_2 \vert \omega_q) =
\Theta(0,\Delta) \, \text{Re} \Big [ f\big ( 4 \omega_q^2 \big) \Big ] - \sum_{n=0}^\infty g_n \mathcal J^{(n)}(t_2, \omega_q, \Delta) \,,
\end{equation}
where $\Theta(0,\Delta) = 1/2$ for the specific choice given in \Eq~\eqref{eq:ThetaDef}. For the second term we have substituted the expansion
\begin{align}
\mathcal G \big ( 2 \omega_q [ 1 + x/(2 \omega_q t_2)] , \, \omega_q \big ) & \equiv \sum_{n=0}^\infty g_n \frac{x^n}{(2 \omega_q t_2)^n} \,,
\end{align}
and have introduced the basis of integrals
\begin{equation}
\mathcal J^{(n)}(t_2, \omega_q, \Delta) \equiv \frac{1}{2 \pi} \frac{1}{(2 \omega_q t_2)^n} \int_{-2 (\omega_q - m_\pi)t_2}^{\infty} dx \, x^n e^{-x } \,
\frac{\Theta( x , t_2 \Delta)}{ 1 + x/( 2 \omega_q t_2) } \mathcal P \frac{1}{x} \,.
\label{eq:Jdef}
\end{equation}
Note here that the principal value is only required for $n=0$. Substituting $g_0 = 2 \, \text{Im} f\big ( 4 \omega_q^2 \big)$, demonstrated in \App~\ref{app:optical}, then gives
\begin{equation}
\label{eq:finaCThetaThree}
c^\Theta_{\,\vec q, -\vec q}(t_2 \vert \omega_q) = \Theta(0,\Delta) \, \text{Re} \Big [ f\big ( 4 \omega_q^2 \big) \Big ] - 2 \, \text{Im} \Big [ f\big ( 4 \omega_q^2 \big) \Big ] \mathcal J^{(0)}(t_2, \omega_q, \Delta) - \sum_{n=1}^\infty g_n \mathcal J^{(n)}(t_2, \omega_q, \Delta) \,.
\end{equation}

\begin{figure}
\begin{center}
\includegraphics[width=0.7\textwidth]{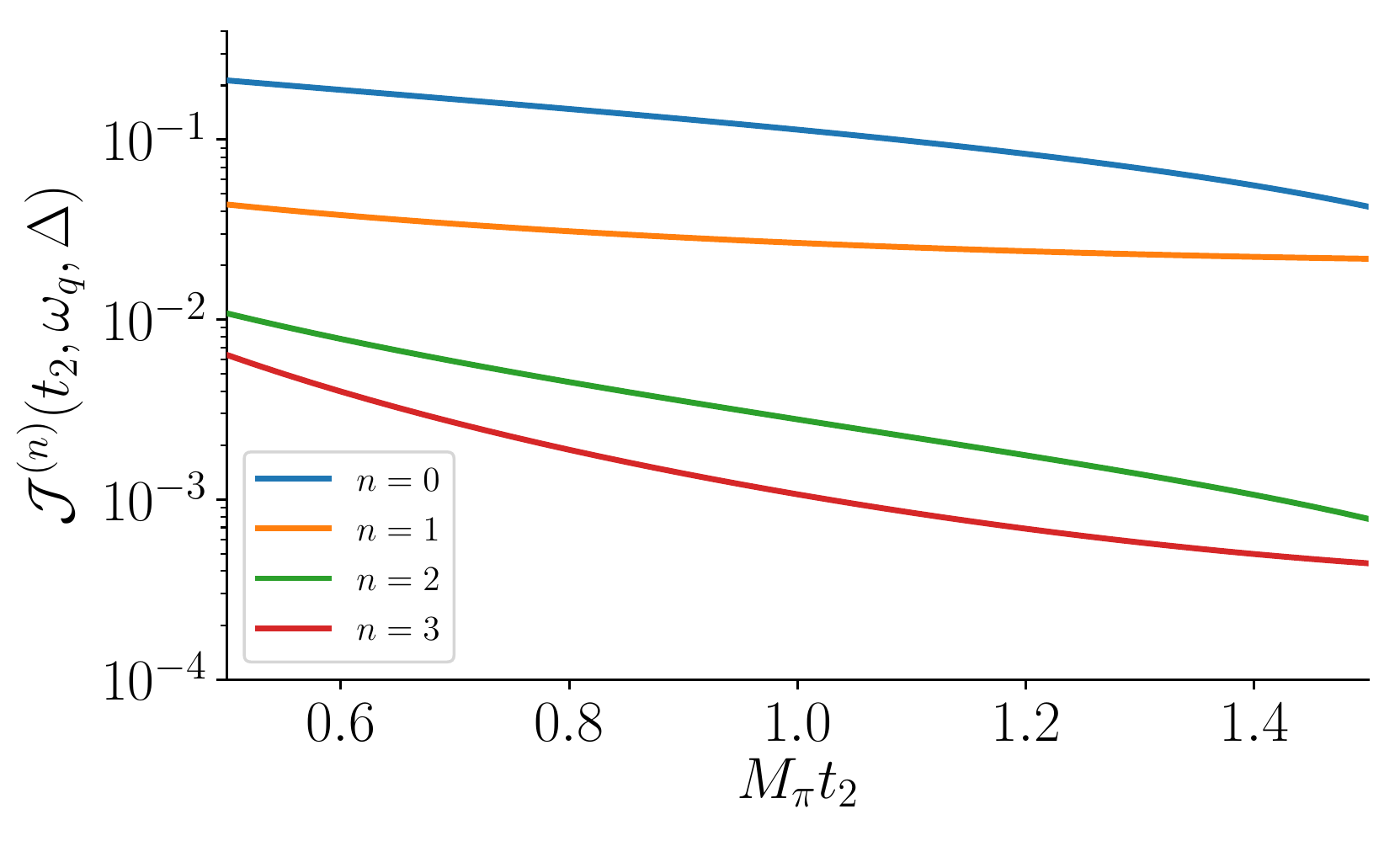}
\caption{Plot of the first few $\mathcal J^{(n)}(t_2, \omega_q, \Delta)$ functions, defined in \Eq~\eqref{eq:Jdef} of the main text, vs $t_2$ with $\omega_q=3m_\pi$ and $\Delta = m_\pi$. \label{fig:Jints}}
\end{center}
\end{figure}

As with the threshold case, the coefficients $g_n$ for $n>1$ contain off-shell derivatives of the scattering amplitude. [See e.g.~\Eq~\eqref{eq:g1Thresh}.] In \Fig~\ref{fig:Jints} we plot the functions $\mathcal J^{(n)}(t_2, \omega_q, \Delta) $ for fixed values of $\omega_q$ and $\Delta$, as a function of $t_2$. We stress that these expressions require $\Delta > 0$ in order for the principal-value pole prescription to properly regulate the integral.

We close this subsection by addressing a final subtlety concerning the fact that the sum over $\mathcal J^{(n)}$ in \Eq~\eqref{eq:finaCThetaThree} [as well as that over $\mathcal I^{(n)}$ in \Eq~\eqref{eq:CThetaThresh}] is, in fact, a divergent asymptotic series. The detailed properties of this series depend on the coefficients $g_n$, but the divergence is expected as $\mathcal J^{(n)} \sim n!$ for large $n$. We stress however that the task here is to estimate the difference between $c^{\Theta}$ and the first two terms on the right-hand side of \Eq~\eqref{eq:finaCThetaThree}. This difference is finite and the divergent asymptotic series provides a representation of it in the usual way, i.e.
\begin{multline}
\label{eq:CThetaAsymp}
c^\Theta_{\,\vec q, -\vec q}(t_2 \vert \omega_q) -
\Theta(0,\Delta) \, \text{Re} \Big [ f\big ( 4 \omega_q^2 \big) \Big ] + 2 \, \text{Im} \Big [ f\big ( 4 \omega_q^2 \big) \Big ] \mathcal J^{(0)}(t_2, \omega_q, \Delta) \\
= - \sum_{n=1}^N g_n \mathcal J^{(n)}(t_2, \omega_q, \Delta) + \mathcal O(\mathcal J^{(N+1)}) \,.
\end{multline}
Given the clear hierarchy in the $\mathcal J^{(n)}$ functions, it should be feasible to keep the first few terms, with the $g_n$ as fit parameters and remove the contamination in order to extract the form factor.


\section{Two-to-two scattering amplitudes\label{sec:4pt}}

We now imitate the results of the previous section, but focusing here on a four-point function built from two nucleon and two pion fields. We begin with the four-point analog of \Eq~\eqref{eq:cThetaDef}, defining
\begin{multline}
\label{eq:cThetaNpiDef}
c^{\Theta, N \pi}_{\vec q_1' \vec q_2' \vec q_1 \vec q_2}(t', t \vert \, M_0) \equiv \frac{ 2 \omega_{\vec q_2'} e^{\omega_{\vec q_2'} t' } 2 \omega_{\vec q_2 } e^{- \omega_{\vec q_2 } t } }{Z_N} C_{a'b'} C_{ab} \, \\
\times \langle \pi^{a'} , \vec q_1' \vert \widetilde N^{b'}_{ \vec q_2'}(0) \, \Theta(\hat M - M_0, \Delta) \, e^{- \hat H(t' - t)} e^{ \omega_{\vec q_1'} t'} e^{ - \omega_{\vec q_1} t} \, \widetilde N^{\dagger b}_{-\vec q_2}(0) \vert \pi^{a}, \vec q_1 \rangle \,.
\end{multline}
Here $a$ and $a'$ are isospin indices for the pions and $b$ and $b'$ are combined spin and isospin indices. The coefficients $C_{ab}$ and $C_{a'b'}$ are chosen to project to $N \pi$ states with definite spin and isospin. As above, here a sum over the repeated indices is implied and the tildes on the operators indicate the projection to definite spatial momentum. Note that momenta with a $1$ subscript correspond to pion fields and those with a $2$ to nucleons, such that
\begin{equation}
\omega_{\vec q_1} = \sqrt{m_\pi^2 + \vec q_1^2} \,, \ \ \ \ \ \ \omega_{\vec q_2} = \sqrt{m_N^2 + \vec q_2^2} \,,
\end{equation}
where $m_N$ is the nucleon mass. Up to the $\Theta$ function, this quantity is directly extractable from a standard Euclidean correlator
\begin{equation}
C_{a'b'} \, \langle \widetilde \pi^{a'}_{\vec q_1'}(t'') \, \widetilde N^{b'}_{\vec q_2'}(t') \,
\widetilde N^{\dagger b}_{-\vec q_2} (t) \, \widetilde \pi^b_{\vec q_1} (0) \rangle \, C_{ab} \,.
\end{equation}
The strategy for implementing $\Theta(\hat M - M_0, \Delta)$ is discussed in the next section.\footnote{Here we denote the inflection point of $\Theta$ by $M_0$ rather than $2 \omega_0$, which is more convenient since the two hadrons are non-degenerate.}

Inserting a complete set of states then yields
\begin{multline}
\label{eq:CThetatoABNpi}
c^{\Theta, N \pi}_{\vec q_1' \vec q_2' \vec q_1 \vec q_2}(t', t \vert \, M_0) = \sum_k \int \! d \Phi_k \, e^{- ( E(\vec p) - \omega_{\vec q_1'} - \omega_{\vec q_2'}) t' } e^{ + ( E(\vec p) - \omega_{\vec q_1 } - \omega_{\vec q_2 }) t } \\[-5pt]
\times \Theta\big ( s(\vec p)^{1/2} - M_0, \Delta \big ) \, A_k(\vec q_1', \vec q_2'; \vec p ) A^*_k(\vec q_1, \vec q_2 ; \vec p ) \,,
\end{multline}
in direct analog to \Eq~\eqref{eq:CThetatoAB} above, here with
\begin{align}
A_k(\vec q_1', \vec q_2'; \vec p ) & \equiv 2 \omega_{\vec q_2'} \frac{C_{a'b'} \, \langle \pi^{a'}, \vec q_1' \vert \, N^{b'} (0) \, \vert k, \mathrm{out}, \vec p_1 \dots \vec p_n \rangle }{\sqrt {Z_N}} \,, \\
A_k(\vec q_1', \vec q_2'; \vec p ) & = \delta_{k,1}
(2 \pi)^3 2 \omega_{\vec q_1'} 2 \omega_{\vec q_2'} \delta^3(\vec p_1 - \vec q_1') - 2 \omega_{\vec q_2'} \frac{\boldsymbol {\mathcal M} _k\big (\eta(\vec q', \vec p) \vert \, \vec \mu(\vec q', \vec p) \big )^{\!*}}{ \eta( \vec q', \vec p) - i \epsilon} \,,
\label{eq:AfinalNpi}
\end{align}
where $k=1$ represents the $N \pi$ channel projected by $C_{ab}$ and only a single delta function arises, since only a single contraction can appear in the disconnected contribution. Note that this decomposition assumes that no two-particle bound state appears in the channel of interest, an assumption we have already noted in footnote 3. If a bound state is present, this can be subtracted from the correlator to reach a new quantity, for which the decomposition here still applies.

Substituting \Eq~\eqref{eq:AfinalNpi} into \Eq~\eqref{eq:CThetatoABNpi} yields four terms: one doubly disconnected, two singly disconnected, and the final term proportional to $\vert \mathcal M \vert^2$. Here we will assume the doubly disconnected piece is subtracted, using a subscript ${\sf c}$ to denote the resulting connected correlator. As we prove in \App~\ref{app:details}, simplifying the remaining terms and setting $\vec q_1 = - \vec q_2 = \vec q$ and $\vec q_1' = - \vec q_2' = \vec q'$ then gives
\begin{align}
\begin{split}
\label{eq:cThetaNpiv3}
c^{\Theta, N \pi}_{\vec q', \vec q}(t', t \vert \, M_0) _{\sf c} =& - e^{- [ \mathcal E(q) - \mathcal E(q') ] t' } \Theta\big ( \mathcal E(q) - M_0, \Delta \big ) \, 2 \omega_{N, q'} \frac{\text{Re} \, \boldsymbol {\mathcal M} \big (\eta( q', q) \vert \, \vec \mu(\vec q', \vec q) \big ) }{ \eta( q', q) } \\
& - e^{- [ \mathcal E(q) - \mathcal E(q')] t } \, \Theta\big ( \mathcal E(q') - M_0, \Delta \big ) \, 2 \omega_{N, q} \frac{\text{Re} \, \boldsymbol {\mathcal M} \big (\eta( q, q') \vert \, \vec \mu(\vec q, \vec q') \big ) }{ \eta( q, q') } \\[5pt]
& + \mathcal C(t', t \vert \vec q', \vec q, M_0, \Delta) \,,
\end{split}
\end{align}
where we have suppressed the $k=1$ index, introduced mass labels on the $\omega$s (since the momenta no longer give a distinction) and also introduced $\mathcal E(q) = \omega_{\pi,q} + \omega_{N,q}$ and
\begin{align}
\eta( q', q) & \equiv \big (\mathcal E(q) - \omega_{\pi, q'} \big )^2 - (\omega_{N,q'})^2 \,, \\
\vec \mu (\vec q',\vec q) & \equiv \big (s, t \big ) = \Big ( \mathcal E(q)^2 \,, \ (\omega_{N,q} - \omega_{N,q'})^2 - (\vec q - \vec q')^2 \Big ) \,.
\end{align}
In \Eq~\eqref{eq:cThetaNpiv3} we have also used the fact that the imaginary parts cancel between the various contributions, as we demonstrate in \App~\ref{app:details}.

In the next paragraph we discuss the final term in \Eq~\eqref{eq:cThetaNpiv3}, $\mathcal C(t', t \vert \vec q', \vec q, M_0, \Delta) $, a direct analog of the last term in \Eq~\eqref{eq:CThetatoABv2} in \Sec~\ref{sec:3pt}. Before doing so, we note here that the first two terms are individually divergent as $q \to q' $. However, these singularities cancel between the two terms such that the final result is finite. Some tedious algebra finds the following result for the $ q \to q' $ limit
\begin{align}
\hspace{-20pt} c^{\Theta, N \pi}_{\vec q}(t', t \vert \, M_0) _{\sf c} & = \text{Re} \, {\mathcal M}\big(\vec \mu(\vec q', \vec q ) \big) \Big [\Theta \big (\mathcal E(q)-M_0,\Delta \big) ( t'-t+1/\omega_{N,q}) - \Theta^{(1,0)} \big ( \mathcal E(q) -M_0 ,\Delta \big) \Big ] \nonumber \\
& \hspace{0pt} - 4 \omega_{N,q} \Theta \big (\mathcal E(q)-M_0,\Delta \big) \Big [ \frac{\mathcal E(q) }{2 \omega_{N,q}} \text{Re} {\mathcal M}^{(1,0)} \big ( \vec \mu(\vec q', \vec q) \big ) + \text{Re} \vec {\mathcal M}^{(1,0,0)} \big (0 \vert \vec \mu(\vec q', \vec q) \big ) \Big ] \nonumber \\[2pt]
& \hspace{0pt} + \mathcal C(t', t \vert \vec q', \vec q, M_0, \Delta) \,,
\label{eq:cThetaNpiFinal}
\end{align}
where the superscript numbers indicate derivatives with respect to the arguments, i.e.~$\eta, s, t$ in the case of $\boldsymbol {\mathcal M} $. This is a striking result: up to the contaminations from the final term, to which we turn shortly, the correlator gives an estimator of the on-shell $N \pi \to N \pi$ scattering amplitude. This is contaminated by an off-shell term with superscript ${}^{(1,0,0)}$, but with a different time dependence, such that the physical result can be disentangled. This result holds for $q = q'$ and $\text{Re} \, \mathcal M\big ( \vec \mu(\vec q, \vec q') \big ) $ simply represents the real part of on-shell amplitude for $N(\vec q) + \pi (- \vec q) \to N(\vec q') + \pi(-\vec q')$.

We turn now to $\mathcal C$, the contamination resulting from the connected parts of both $A_k$ and $A^*_k$. For the simplifying case of $ q' = q $, the fully connected quantity is given by
\begin{multline}
\label{eq:calC4ptdef}
\mathcal C(t', t \vert \vec q', \vec q, M_0, \Delta) = \int_{m_\pi + m_N}^\infty \! \frac{d \mathcal E}{2 \pi} \, e^{- ( \mathcal E - \mathcal E(q)) (t' -t) } \Theta( \mathcal E - M_0, \Delta) \\[-2pt] \times
\, \mathcal H(\mathcal E, \vec q', \vec q) \, \frac{(2 \omega_{N,q})^2}{ ( \mathcal E- \omega_{\pi, q} + \omega_{N,q} )^2}
\bigg [ \mathcal P \frac{ 1}{ \mathcal E - \mathcal E(q)} \bigg ]^2 \,,
\end{multline}
where
\begin{align}
\label{eq:calHdef}
\mathcal H(\mathcal E, \vec q', \vec q) & = \sum_k \int d \Phi_k \, 2 \pi \delta \big( \mathcal E - E(\vec p) \big ) \, \boldsymbol {\mathcal M} _k\big (\eta(\vec q, \vec p) \vert \, \vec \mu(\vec q, \vec p) \big )^{\!*} \boldsymbol {\mathcal M} _k\big (\eta(\vec q', \vec p) \vert \, \vec \mu(\vec q', \vec p) \big ) \,.
\end{align}
Applying the change of variables $x = ( \mathcal E - \mathcal E(q)) (t' -t)$ then yields
\begin{equation}
\label{eq:calC4ptFinal}
\mathcal C(t', t \vert \vec q', \vec q, M_0, \Delta) = \int_{- ( \mathcal E(q) - m_\pi - m_N)(t'-t)}^\infty \! \! \! dx \, \mathcal H\big ( x/(t'-t) + \mathcal E(q), \vec q', \vec q \big ) \, \kappa(x, t' - t) \,,
\end{equation}
where
\begin{equation}
\label{eq:kappa4pt}
\kappa(x, t' - t) = \frac{(t' - t)}{2 \pi } e^{- x } \, \Theta \big ( x/(t'-t) + \mathcal E(q) - M_0, \Delta \big ) \frac{(2 \omega_{N,q})^2}{[ x/(t'-t) + 2 \omega_{N,q} ]^2} \bigg [ \mathcal P \frac{1}{x} \bigg ]^2 \,.
\end{equation}

Equations \eqref{eq:cThetaNpiFinal}, \eqref{eq:calHdef} \eqref{eq:calC4ptFinal}, and \eqref{eq:kappa4pt} summarize the main new technical results of this section. In the following subsections we explore their implications for extracting scattering information.  In direct analogy to our treatment in \Sec~\ref{sec:3pt}, we divide the discussion into threshold and general kinematics.

\subsection{Threshold kinematics} \label{sec:Threshold4pt}

For $\mathcal E$ near threshold and $\vec q = \vec q' = \vec 0$, the function $\mathcal H$ takes the form
\begin{equation}
\mathcal H(\mathcal E, \vec 0, \vec 0) = \int \! \! \frac{d^3 \vec p}{ 2 \omega_{N, \vec p} (2 \pi)^3 } \frac{1}{ 2 \omega_{\pi, \vec p} } \, 2 \pi \delta \big( \mathcal E - \omega_{ \pi , \vec p} - \omega_{ N , \vec p}\big )
\vert \boldsymbol {\mathcal M} _1\big (\eta(\vec 0, \vec p) \vert \, \vec \mu(\vec 0, \vec p) \big ) \vert^2 \,,
\end{equation}
where the coordinates in the scattering amplitude are given explicitly by
\begin{align}
\eta(\vec 0, \vec p) & = (\omega_{ \pi , \vec p} +\omega_{ N , \vec p}- m_\pi )^2 - m_N^2 \,, \\
\vec \mu(\vec 0, \vec p) & = (s,t) = \Big ( ( \omega_{ \pi , \vec p} +\omega_{ N , \vec p})^2 \,, \ (\omega_{\pi, \vec p} - m_\pi)^2 - \vec p^2 \Big ) \,.
\end{align}
Note here that, although the angular dependence is absent for $\vec q = \vec q' = \vec 0$, the amplitude nonetheless depends on the Mandelstam variable $t$ as an artifact of the off-shell kinematics.
Next, we use the fact that the Dirac delta function sets $\vec p^2 = p^2 = k_{ N \pi}(\mathcal E)^2$, with the magnitude of back-to-back momenta for non-degenerate particles given by
\begin{equation}
k_{N \pi}(\mathcal E)^2 = \frac{\mathcal E^2}{4} - \frac{m_\pi^2 + m_N^2}{2} + \frac{(m_N^2 - m_\pi^2)^2}{ 4 \mathcal E^2 } \,.
\end{equation}
Introducing the shorthand
\begin{equation}
\boldsymbol {\mathcal M}_{N \pi}(\mathcal E) = \boldsymbol {\mathcal M} _1\big (\eta(\vec 0, \vec p) \vert \, \vec \mu(\vec 0, \vec p) \big ) \bigg \vert_{\vec p^2 = k_{N \pi}(\mathcal E)^2} \,,
\end{equation}
and evaluating the remaining phase-space integral, we reach
\begin{equation}
\mathcal H(\mathcal E, \vec 0, \vec 0) = \big \vert \boldsymbol {\mathcal M}_{N \pi}(\mathcal E) \big \vert^2 \frac{k_{N \pi}(\mathcal E)}{4 \pi \mathcal E} \,.
\end{equation}

Substituting these simplifications into the definition of $\mathcal C$, the four-point function reduces to
\begin{multline}
\hspace{-10pt} c^{\Theta, N \pi}_{\vec 0}(t', t \vert \, M_0) _{\sf c} =
8 \pi (m_N + m_\pi) \, a_{N\pi} \, (t' - t + 1/m_N) - 2 (m_N + m_\pi) \text{Re} \, \partial_s {\mathcal M}_{\sf s}(s) - 4 m_N \text{Re} \, \partial_{\eta} \vec {\mathcal M}_{\sf s}(\eta \vert s) \\[5pt]
+ \frac{1}{2 \pi } \int_{0}^\infty \! \! \! dx \,
\, \frac{(t' - t) e^{- x }}{[1 + x/[ 2 m_N (t'-t) ] ]^2} \bigg [ \mathcal P \frac{1}{x} \bigg ]^2
\big \vert \boldsymbol {\mathcal M}_{N \pi}(\mathcal E) \big \vert^2 \frac{k_{N \pi}(\mathcal E)}{4 \pi \mathcal E} \bigg \vert_{\mathcal E = x/(t'-t) + m_N + m_\pi} \,.
\label{eq:4ptThreshPenultima}
\end{multline}
In this result, the top line depends only on $\boldsymbol {\mathcal M}_{\sf s}$, the off-shell scattering amplitude projected to zero orbital angular momentum. This differs from $\boldsymbol {\mathcal M}_{N \pi}$ because, for terms that are linear in the scattering amplitude, both the incoming and outgoing spatial momenta are set to zero. As a result the Mandelstam $t$ is set identically to zero and no higher partial waves can contribute. Here we have also used the relation between the threshold scattering amplitude and scattering length for the case of non-degenerate particles
\begin{equation}
\mathcal M_{\sf s}\big ( (m_N + m_\pi)^2 \big ) = 8 \pi (m_N + m_\pi) a_{N \pi} \,.
\end{equation}

In direct analog to our analysis of the three-point function, the final step is to expand the final term of \Eq~\eqref{eq:4ptThreshPenultima}. We reach
\begin{multline}
c^{\Theta, N \pi}_{\vec 0}(t', t \vert \, M_0) _{\sf c} =
8 \pi (m_N + m_\pi) \, a_{N\pi} \, (t' - t + 1/m_N) - 2 (m_N + m_\pi) \text{Re} \, \partial_s {\mathcal M}_{\sf s}(s) \\[5pt]
- 4 m_N \text{Re} \, \partial_{\eta} \vec {\mathcal M}_{\sf s}(\eta \vert s) +
\frac{4(t' - t) }{(1 + m_\pi/m_N)^2} \sum_{n=0}^\infty h_n \, \mathcal K^{(n)}\big ( b , \delta m \big ) \,,
\end{multline}
where in the final term we have introduced $b=(m_N+m_\pi)(t'-t)$ and $\delta m = (m_N-m_\pi)/(m_N+m_\pi)$. We have also introduced
\begin{align}
\mathcal K^{(n)} (b, \delta m) & \equiv \frac{1 }{16 \pi^2 b^{n+1/2}} \int_{ 0}^\infty \! dx \, x^n \, e^{- x } \sqrt{\frac{x (x/b+2)(x/b+1-\delta m)}{(x/b+1+ \delta m)^3 (x/b+1)^4}} \bigg [ \mathcal P \frac{1}{x} \bigg ]^2
\,, \\[5pt]
& = \frac{\sqrt 2}{16 \pi^2 b^{n+1/2}} \Gamma(n-1/2) \frac{(1-\delta m)^{1/2} }{(1+\delta m)^{3/2}} \Big [ 1 + \mathcal O \big ( 1/b \big ) \Big ] \,, \label{eq:Kexp} \\[5pt]
h_n & \equiv \frac{1}{n!} \ (m_N+m_\pi)^n \frac{d^n}{d\mathcal E^n} \big \vert \boldsymbol {\mathcal M}_{N \pi} ( \mathcal E ) \big \vert^2_{\mathcal E = m_N + m_\pi } \,.
\end{align}

In the case of degenerate, non-identical particles the kernel simplifies to
\begin{align}
\label{eq:KDEGENdef}
\mathcal K^{(n)} (b, 0) & \equiv \frac{1 }{16 \pi^2 b^{n+1/2}} \int_{ 0}^\infty \! dx \, x^n \, e^{- x } \frac{ \sqrt{x (x/b+2)}}{(x/b+1)^3} \bigg [ \mathcal P \frac{1}{x} \bigg ]^2 \,, \\[5pt]
& = \frac{\sqrt 2}{16 \pi^2 b^{n+1/2}} \Gamma(n-1/2) \Big [ 1 + \mathcal O \big ( 1/b \big ) \Big ] \label{eq:Kexp} \,,
\end{align}
and in the case of identical particles one requires an additional factor of $(1/2)$ in these expressions.

\begin{figure}
\begin{center}
\includegraphics[width=0.7\textwidth]{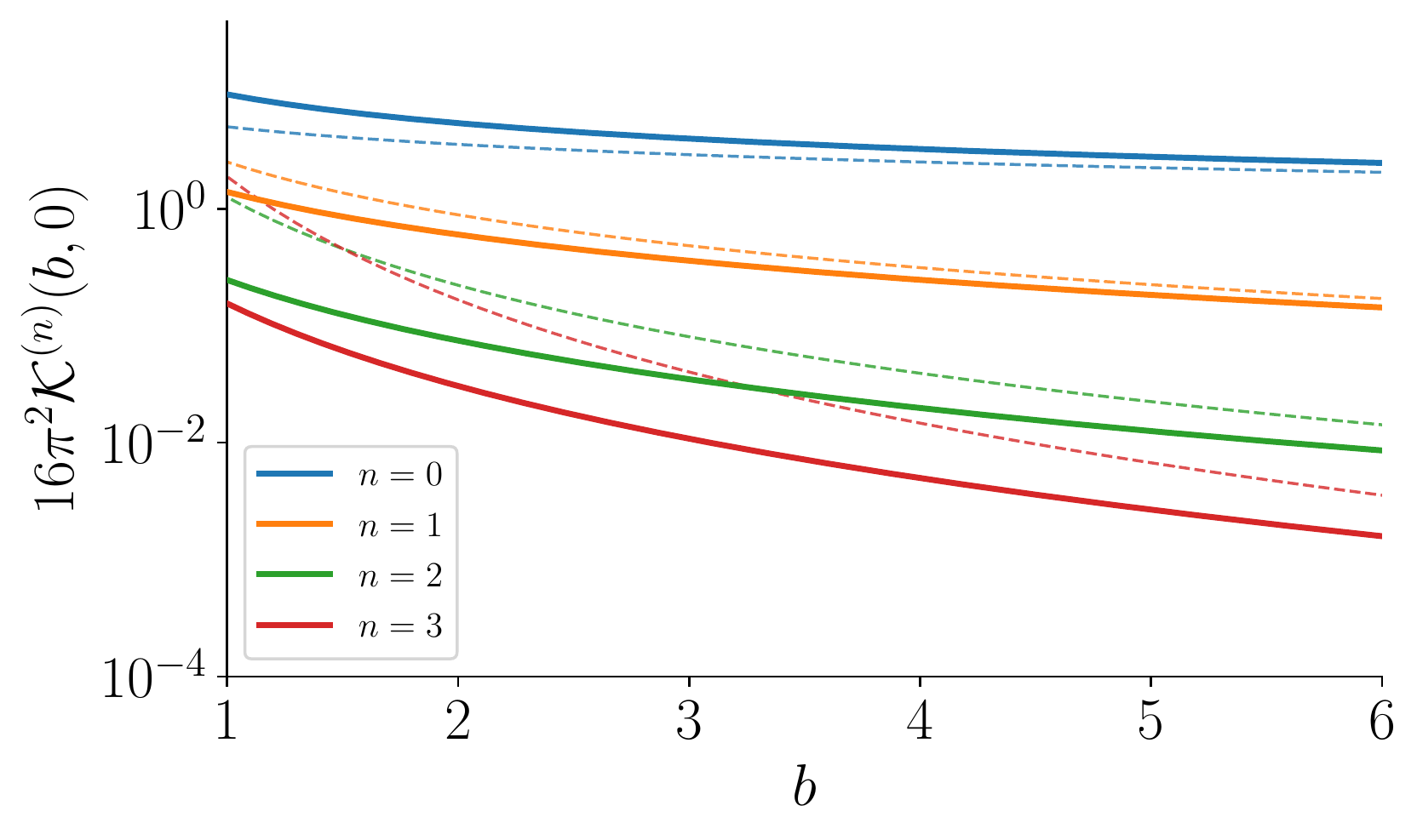}
\caption{Plot of the first few $\mathcal K^{(n)}(b,0)$ functions (solid lines), defined in \Eq~\eqref{eq:KDEGENdef} of the main text. The dashed lines show the large $b$ expansion for each function, scaling as $b^{-n-1/2}$ as shown in \Eq~\eqref{eq:Kexp}. \label{fig:Kints}}
\end{center}
\end{figure}

Expanding $\mathcal K^{(0)}(b, \delta m)$ in powers of $t'-t$ gives the direct analog of \Reference~\cite{Maiani:1990ca} for the four-point function
\begin{multline}
c^{\Theta, N \pi}_{\vec 0}(t', t \vert \, M_0) _{\sf c} = 8 \pi (m_N + m_\pi) \, a_{N\pi} \, (t' - t ) \\
- 16 \, a_{N \pi}^2 \sqrt{2 \pi (m_N + m_\pi) m_N m_\pi (t'-t)} + \mathcal O \big ((t'-t)^0\big) \,.
\end{multline}
This was already highlighted in the introduction, see \Eq~\eqref{eq:NpiThreshIntro}. The relation appears promising as two separate time dependences, together with $a_{N\pi}$ and $a^2_{N\pi}$ coefficients, may give a strong constraint of this threshold observable. As with the three-point function, one reaches a better description by working in the basis of $\mathcal K$ integrals. See \Fig~\ref{fig:Kints}.

\subsection{General kinematics} \label{sec:General4pt}

Finally, the general-energy result for the four-point function, with $M_0 = \mathcal E(q)$, reads
\begin{align}
\begin{split}
\hspace{-20pt} c^{\Theta, N \pi}_{\vec q}(t', t \vert \, M_0) _{\sf c} & = \text{Re} \, {\mathcal M}\big(\vec \mu(\vec q', \vec q ) \big) \Big [ \frac{ t'-t+1/\omega_{N,q}}{2} - \frac{1}{2 \Delta} \Big ] \\[5pt]
& \hspace{0pt} -\mathcal E(q) \text{Re} {\mathcal M}^{(1,0)} \big ( \vec \mu(\vec q', \vec q) \big )- 2 \omega_{N,q} \text{Re} \vec {\mathcal M}^{(1,0,0)} \big (0 \vert \vec \mu(\vec q', \vec q) \big ) \\[3pt]
& \hspace{0pt} + (t' - t) \sum_{n=0}^\infty h_n \mathcal L^{(n)}(t'-t, q, \Delta) \,,
\end{split}
\end{align}
where we have used $\Theta(0,\Delta) = 1/2$ and also substituted the derivative of the theta function. The final term is built from the integrals
\begin{multline}
\label{eq:Ldef}
\mathcal L^{(n)}(t'-t, q, \Delta) \equiv \frac{1}{2 \pi } \frac{1}{[ \mathcal E(q) (t'-t)]^n} \int_{- (\mathcal E(q) - m_\pi - m_N) (t'-t)}^{\infty} \! \! \! \! dx \, x^ne^{- x }
\\ \times \frac{ \Theta \big ( x , \Delta (t'-t) \big ) }{ \big ( 1+ x/[ 2 \omega_{N,q} (t'-t)] \big )^2} \bigg [ \mathcal P \frac{1}{x} \bigg ]^2 \,,
\end{multline}
together with the coefficients $h_n$, defined via
\begin{align}
\mathcal H\big ( x/(t'-t) + \mathcal E(q), \vec q', \vec q \big ) & \equiv \sum_{n=0}^\infty h_n \frac{x^n}{[ \mathcal E(q) (t'-t)]^n} \,.
\end{align}
Substituting $h_0 = \text{Im} \, \mathcal M\big ( \vec \mu(\vec q, \vec q') \big )$, demonstrated in \App~\ref{app:details}, then gives
\begin{align}
\begin{split}
c^{\Theta, N \pi}_{\vec q}(t', t \vert \, \omega_q) _{\sf c} & = \text{Re} \, {\mathcal M}\big(\vec \mu(\vec q', \vec q ) \big) \Big [ \frac{ t'-t+1/\omega_{N,q}}{2} - \frac{1}{2 \Delta} \Big ] \\[3pt]
& + \text{Im} \, \mathcal M\big ( \vec \mu(\vec q, \vec q') \big ) \, (t'-t) \, \mathcal L^{(0)}(t'-t, q, \Delta) \\[7pt]
& - \mathcal E(q) \text{Re} {\mathcal M}^{(1,0)} \big ( \vec \mu(\vec q', \vec q) \big )- 2 \omega_{N,q} \text{Re} \vec {\mathcal M}^{(1,0,0)} \big (0 \vert \vec \mu(\vec q', \vec q) \big ) \\[0pt]
& + (t' - t) \sum_{n=1}^\infty h_n \mathcal L^{(n)}(t'-t, q, \Delta) \,.
\label{eq:4ptGenKinFin}
\end{split}
\end{align}
The first two lines represent physical information and the final two represent the contaminations that one must remove via fitting.

This completes our presentation of strategies for extracting scattering information from Euclidean correlators, at both threshold and general kinematics. We now turn to discussing how the modified correlation function, defined with a smooth $\Theta$ function inserted with the interpolating operators, might be extracted in practice, in a numerical lattice calculation.

\begin{figure}
\begin{center}
\includegraphics[width=0.7\textwidth]{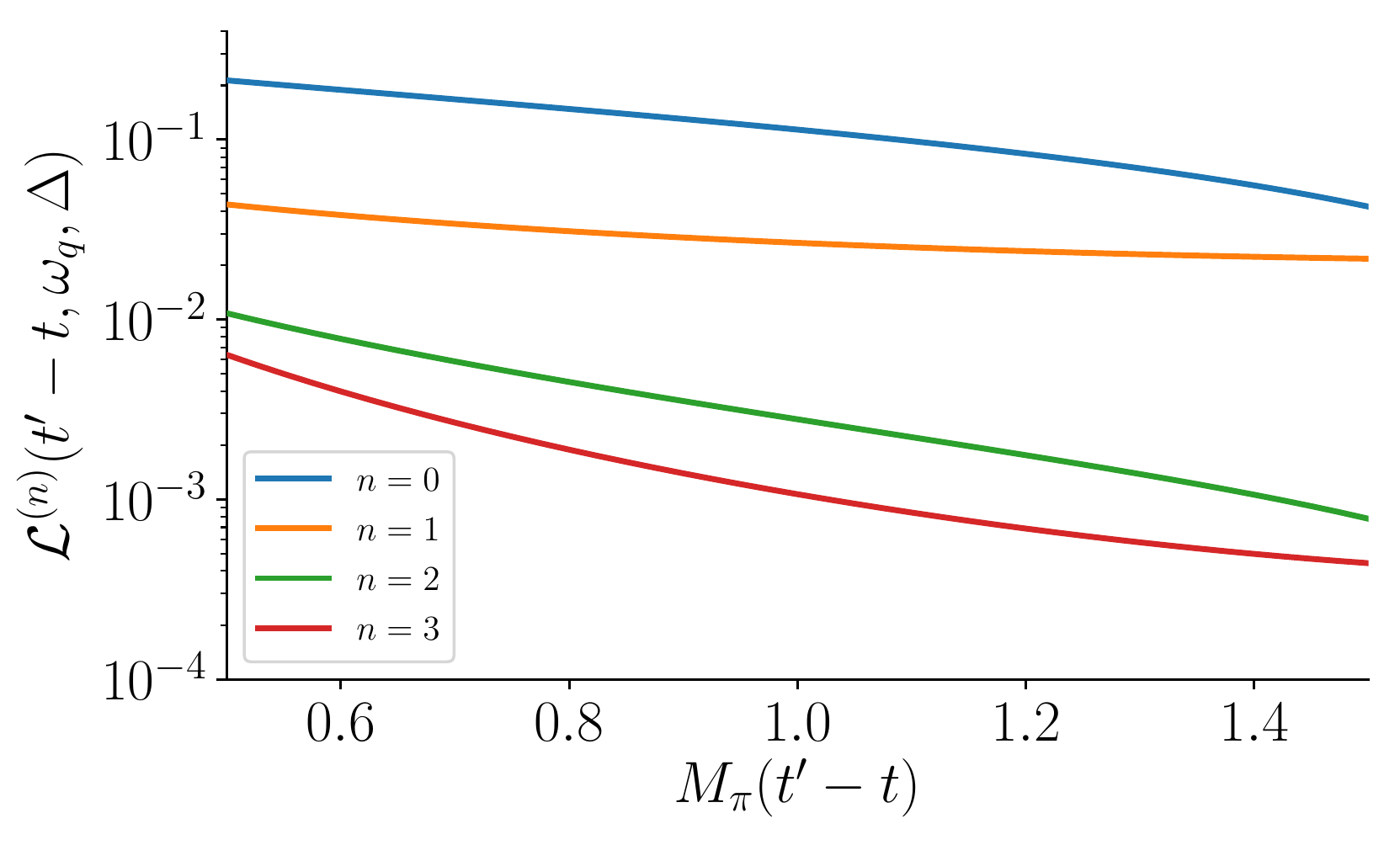}
\caption{Plot of the first few $\mathcal L^{(n)}(t'-t, \omega_q, \Delta)$ functions, defined in \Eq~\eqref{eq:Ldef} of the main text, vs $(t'-t)$. Here we consider the functions in the case of two degenerate particles with mass $m_\pi$ and with $\omega_q=3m_\pi$ and $\Delta = m_\pi$. \label{fig:Lints}}
\end{center}
\end{figure}


\section{Implementation strategies\label{sec:imp}}

In this section we discuss two possibilities to estimate the modified quantities $c^\Theta$ and $c^{\Theta,N\pi}$, defined in \Eqs~\eqref{eq:cThetaDef} and \eqref{eq:cThetaNpiDef} respectively. We make one adjustment relative to the previous section by replacing the $N \pi \to N \pi$ four-point function with the simplified version built only out of pion fields. We denote the latter by $c^{\Theta,\pi \pi}$. With straightforward adjustments of the previous section, in particular setting $\delta m = 0$ and including the $(1/2)$ for identical particles, this can be used to extract the $\pi \pi \to \pi \pi$ scattering amplitude.

In the previous sections, we have not addressed the fact that numerical lattice calculations are necessarily confined to a finite space-time volume. In this section we consider the effects of a finite cubic spatial volume, with periodicity $L$ in each of the three spatial directions, but continue to neglect the thermal effects of a finite temporal extent. Within this set-up, suppose that one has extracted the matrix element
\begin{equation}
M_3^i (t) \equiv \langle \pi, \vec q_i , L | \widetilde \pi_{-\vec q_i} (t) \widetilde J_{\vec 0}(0) \vert 0 \rangle \,,
\label{eq:M3def}
\end{equation}
from the relevant three-point function. Here the index $i$ labels the values of non-equivalent back-to-back momenta. In practice the combined pion state and pion field should be projected to a definite two-pion isospin along the lines of the $N \pi$ projections involving the $C_{ab}$ coefficients. For the scalar current, $\widetilde J_{\vec 0}(t) = \int d^3 \vec x \, J(x)$, the total isospin must be $0$ or $2$ to give a non-zero result.

Similarly, we consider the two-point function and also matrix elements from four-point functions:
\begin{equation}
M_2(t) \equiv \langle \widetilde J_{\vec 0}(t) \, \widetilde J_{\vec 0}(0) \rangle\,, \quad
M_4^{ij} (t) \equiv \langle \pi, \vec q_i , L | \, \widetilde \pi_{-\vec q_i} (t) \,
\widetilde \pi_{\vec q_j} (0) \, | \pi, \vec q_j , L \rangle \,.
\end{equation}
It is then useful to assemble these quantities into a matrix of correlation functions
\begin{equation}
M(t) \equiv \begin{pmatrix}
M_2(t) & M_3^T(t) \\
M_3(t) & M_4(t) \\
\end{pmatrix} \,,
\end{equation}
with rank dictated by the number of pion momenta available in the calculation. Following \References~\cite{WilsonGEVP,Blossier:2009kd}, from the solution of the generalized eigenvalue problem on this matrix, one can extract a series of finite-volume energies, $E_n(L)$, and matrix elements
\begin{align}
M_3^i[n] & = \langle \pi, \vec q_i, L | \widetilde \pi_{- \vec q_i}(0) | n, L \rangle \,
\langle n, L | \widetilde J_{\vec 0}(0) | 0 \rangle \,, \\
M_4^{ij}[n] & = \langle \pi, \vec q_i , L | \widetilde \pi_{- \vec q_i} (0) | n, L \rangle \,
\langle n, L | \widetilde \pi_{ \vec q_j} (0) | \pi, \vec q_j, L \rangle \,,
\end{align}
up to a maximal value of $E_N(L)$, depending on the size of the operator basis.

Then the matrix element used in Eq.~\eqref{eq:cThetaDef} to define the modified quantity $c^\Theta_{\vec q, - \vec q}(t_2 | \omega_q)$ can be constructed as
\begin{multline}
c^{\Theta}_{\vec q_i, - \vec q_i}(t_2 \vert \omega_0) \equiv \frac{ 2 \omega_{ q_i} e^{\omega_{ q_i} t_2 } }{\sqrt{Z_\pi}} \times \\[-5pt]
\bigg [ \langle \pi, \vec q_i, L | \widetilde \pi_{-\vec q_i} (t_2) \widetilde J_{\vec 0}(0) | 0 \rangle - \sum_{n=1}^N \Theta \big (2 \omega_{q_i} - E_n(L), \Delta \big ) \, M_3^i[n] \, e^{-(E_n(L) - \omega_{ q_i}) t_2} \bigg ] \,,
\end{multline}
provided that the given value of $2\omega_{q_i}$ is smaller than $E_N(L)$ to avoid systematic errors due to the truncated sum.

The additional effort in measuring the various correlation functions required to construct $M(t)$ is compensated by the fact that we can study several kinematic points for the $2 \to 1$ transitions, together with the $2 \to 2$ amplitudes
\begin{multline}
\langle \pi, \vec q_i | \widetilde \pi_{-\vec q_i} (0) \,
\Theta(\hat M - 2 \omega_{q_i}, \Delta) e^{-(\hat H - \omega_{\vec q_i}) t} \,
\widetilde \pi_{\vec q_j} (0) | \pi, \vec q_j \rangle = \\
\langle \pi, \vec q_i | \widetilde \pi_{-\vec q_i} (t)
\widetilde \pi_{\vec q_j} (0) | \pi, \vec q_j \rangle - \sum_{n=1}^N \Theta \big (2 \omega_{q_i} - E_n(L), \Delta \big ) \,
M_4^{ij}[n] \,
e^{-(E_n(L) - \omega_{ q_i}) t} \,.
\end{multline}

Note that this procedure in fact relies on finite-$L$, in order to ensure that a finite (and manageable) number of states appear in the regime where $\Theta \big (2 \omega_{q_i} - E_n(L), \Delta \big )$ has support. Once the re-construction is achieved, the resulting $c^\Theta$ and $c^{\Theta,\pi\pi}$ are expected to have residual volume effects, scaling as $\mathcal O(e^{- L \Delta})$. Thus, exactly as with the methods described in \References~\cite{Hansen:2017mnd,Bulava:2019kbi}, it is important that $ L \Delta$ is sufficiently large that these can be removed or included in the systematic uncertainty. Future work, perhaps along the lines of \References~\cite{Hansen:2019rbh,Hansen:2020whp} is needed to fully analyze the finite-$L$ corrections to these correlators as a function of the $\Theta$ width as well as the time and energy coordinates.

Alternatively, for situations where an exclusive study is not possible, e.g.~for higher center-of-mass energies where a variational approach would be impractical, approximate solutions to the inverse problem represent a viable alternative. The approach is to define coefficients, $w^{\Theta}(t, \Delta, \omega_{q_i} \vert t')$, satisfying
\begin{equation}
\sum_{t'} \, w^{\Theta}(t, \Delta, \omega_{q_i} \vert t') \, e^{- (\mathcal E - \omega_{q_i}) t'} = \Theta( \mathcal E - 2 \omega_{q_i} ,\Delta) \, e^{-(\mathcal E - \omega_{q_i}) t} \,,
\end{equation}
where the kernels $w$ are found numerically, for each value of $t$, $\Delta$ and $\omega_{q_i}$, using Backus-Gilbert-like methods~\cite{Backus,Hansen:2019idp}, or Chebyshev polynomials~\cite{Bailas:2020qmv}. These then satisfy
\begin{multline}
\langle \pi, \vec q_i | \widetilde \pi_{-\vec q_i} (0) \,
\Theta(\hat M - 2 \omega_{q_i}, \Delta) e^{-(\hat H - \omega_{\vec q_i}) t} \,
\widetilde \pi_{\vec q_j} (0) | \pi, \vec q_j \rangle = \\[3pt]
\sum_{t'} \, w^{\Theta}(t, \Delta, \omega_{q_i} \vert t') \,
\langle \pi, \vec q_i | \widetilde \pi_{-\vec q_i} (t')
\widetilde \pi_{\vec q_j} (0) | \pi, \vec q_j \rangle \,.
\end{multline}
While a large $\Delta$ improves the stability of the numerical solutions, the change of basis from $e^{-Et}$ to $\Theta(E-2\omega_{q_i},\Delta) e^{-Et}$ may also simplify the complexity of the inverse problem (compared to a target Gaussian or Breit-Wigner peak, for example) thereby potentially reducing systematic errors.

Finally, we note that it may also be useful to construct the coefficients to reproduce $1 - \Theta( \cdots ) $ instead, i.e.~the reflected $\Theta$-function that decreases with increasing $\mathcal E$. The resulting estimator can then be subtracted from the original correlator to reach the desired quantity. This seems like a trivial adjustment as the two sets of coefficients should formally be related as
\begin{equation}
w^{\Theta}(t, \Delta, \omega_{q_i} \vert t') = \delta_{t,t'} - w^{1 - \Theta}(t, \Delta, \omega_{q_i} \vert t') \,.
\end{equation}
In practice, however, the algorithms used to determine the coefficients depend non-trivially on the target functions and $w^{1 - \Theta}$ may give a better overall estimate.


\section{Summary and outlook}

In this article we have considered extensions of the work of Maiani and Testa \cite{Maiani:1990ca}. These divide into two main categories: \emph{(i)} generalizing the results for threshold correlators and \emph{(ii)} proposing modified correlators that allow one to extend the reach of scattering extractions above threshold.

In the case of threshold kinematics, the key results are presented in \Secs~\ref{sec:Threshold3pt} and \ref{sec:Threshold4pt}. In the former we review the analysis of the $\pi \pi J$ three-point function (with $J$ a scalar current and $\pi$ a single-particle interpolator) and consider an alternative expansion in kinematic functions that replaces the original expansion of \Reference~\cite{Maiani:1990ca}, in powers of inverse time ($1/t$), and appears to exhibit better convergence at small to moderate $t$ values. We also give explicit expressions for the leading off-shellness contaminating the correlator. In section~\ref{sec:Threshold4pt}, we extend these considerations to four-point functions, focusing on $N \pi \to N \pi$. This gives a tool to extract the $N \pi$ scattering length, $a_{N\pi}$, from a non-standard fit to the Euclidean time dependence, with terms scaling as $ a_{N \pi} t$ and $ a_{N \pi}^2 \sqrt{t}$; see also \Eq~\eqref{eq:NpiThreshIntro} in the introduction.

The second class of extensions, targeting scattering information away from threshold, are detailed in \Secs~\ref{sec:General3pt} and \ref{sec:General4pt}, again with a focus on three- and four-point functions, respectively. The key idea is to consider a modified correlator, in which a smooth step function, denoted by $\Theta(\hat M - M_0,\Delta)$, is inserted with the operators. Here $\Delta$ denotes the width of the step, $\hat M$ is the hamiltonian boosted to zero-momentum, and $M_0$ a free parameter. This modification effectively shifts the threshold to higher energies, so that the Maiani-Testa approach can be used to extract the form factor and scattering amplitude for general kinematics. As we stress in the main text, this method is analogous to those based in reconstructing the spectral function \cite{Hansen:2017mnd,Bulava:2019kbi}. The key distinction here is that the target energy range is isolated by the combined effect of the $\Theta$-function and the Euclidean-time translation operator, $e^{- \hat H t}$. In particular, the latter naturally provides the damping of contributions above the target energy, see also \Fig~\ref{MTvsINV}. This is formalized by the large time expansions presented in \Secs~\ref{sec:General3pt} and \ref{sec:General4pt}. At leading order these give the real and imaginary parts of the physical observables, separated by coefficients with different time dependencies.

In \Sec~\ref{sec:imp} we briefly describe the implementation strategies for extracting the $\Theta$-modified correlator from numerical lattice calculations. We outline two methods there, one based on the generalized eigenvalue problem (GEVP) and the other on reconstruction methods such as that due to Backus and Gilbert \cite{Backus}. In both cases, this approach may have an advantage over the other methods because only the low-energy modes of the correlator need to be modified. In particular, the contamination of higher energies is encoded in the large $t$ expansion of the correlation function and therefore profits from knowledge of the functional form (ultimately arising from Lorentz invariance and the form of the time-translation operator and threshold singularities). This information is not directly used, for example, in \References~\cite{Hansen:2017mnd,Bulava:2019kbi}. In \Sec~\ref{sec:imp} we also briefly discuss the role of finite volume, and the importance of $\Delta$ in suppressing volume effects.

Looking forward, the next steps are to test these methods in a lattice calculation. The most likely initial applications might include extracting scattering lengths from standard three- and four-point functions, especially on lattice ensembles where this information is available from a finite-volume analysis and can provide a direct comparison. This can be used to test, for example, whether one can achieve a more precise determination due to the fact that the observable appears at leading order in the matrix element, rather than as a shift to the energies. This will set the stage for the more ambitious implementation away from threshold, with ultimate target observables including electroweak decay amplitudes, QED corrections to hadronic matrix elements and long-range matrix elements. This will require some extensions of the formal results presented here, but these are expected to be relatively straightforward, along the lines of \Reference~\cite{Bulava:2019kbi}.

\acknowledgments
We warmly thank
John Bulava,
Marco C\'e,
Anthony Francis,
Dorota Grabowska,
Jeremy Green,
Taku Izubuchi,
Christoph Lehner,
Guido Martinelli,
Harvey Meyer,
Daniel Robaina,
Steve Sharpe
and
Chris Sachrajda
for useful discussions, and for previous and ongoing collaborations that helped to inspire this work. MH was supported by UK Research and Innovation Future Leader Fellowship MR/T019956/1. This work is dedicated to Tommaso Sergio Bruno, who provided invaluable motivation in pushing the manuscript to completion.

\appendix


\section{Technical details}

\subsection{Matrix element decomposition\label{app:ABproof}}

In this section we prove \Eq~\eqref{eq:Afinal} of the main text. Begin with a Minkowski-signature representation of $A_k(\vec q_1, \vec q_2; \vec p)$
\begin{equation}
(2 \pi)^4 \delta^4(q_1+q_2- P) A_k(\vec q_1, \vec q_2; \vec p) = \frac{2 \omega_{\vec q_2} }{\sqrt{Z_\pi}} \int d^4 x \, e^{i q_2^0 x^0 - i \vec q_2 \cdot \vec x} \, \langle \pi, \vec q_1 \vert \pi(x) \vert k, \text{out}, \vec p_1 \cdots \vec p_{n_k} \rangle \,,
\end{equation}
where we have introduced the four-vectors $q_1 = (\omega_{\vec q_1}, \vec q_1), q_2 = (q_2^0, \vec q_2)$ and
\begin{equation}
P \equiv \big ( E(\vec p), \vec P \big ) = \big ( \omega_{\vec p_1} + \cdots + \omega_{\vec p_{n_k}}, \, \vec p_1 + \cdots + \vec p_{n_k} \big ) \,.
\end{equation}
Next we decompose the matrix element on the right-hand side into disconnected and connected components to reach
\begin{multline}
(2 \pi)^4 \delta^4(q_1+q_2- P) A_k(\vec q_1, \vec q_2; \vec p) =\\[5pt]
\hspace{80pt} \delta_{k,1} (2 \pi)^7 2 \omega_{\vec q_1} 2 \omega_{\vec q_2} \big [ \delta^3(\vec p_1 - \vec q_1) \delta^4(p_2 - q_2) + \delta^3(\vec p_2 - \vec q_1) \delta^4(p_1-q_2) \big ] \\[5pt] + (2 \pi)^4 \delta^4(q_1 + q_2 - P)
2 \omega_{\vec q_2} \frac{\boldsymbol {\mathcal M}_k\big (\eta(\vec q, \vec p) \vert \, \vec \mu(\vec q, \vec p) \big )^{\!*}}{ \eta( \vec q, \vec p) - i \epsilon} \,.
\label{eq:AexpressionAPP}
\end{multline}
This result serves to define the off-shell amplitude $\boldsymbol {\mathcal M}_k$ as the connected component of the matrix element with a simple pole factor removed. The latter is defined using
\begin{align}
\eta (\vec q, \vec p) & = ( E(\vec p)- \omega_{\vec q_1})^2 - (\vec P - \vec q_1)^2 - m_\pi^2 \,.
\label{eq:etazetadefAPP}
\end{align}
Note that the self-energy contributions from the external leg are absorbed into the off-shell definition.

From the Lehmann-Symanzik-Zimmermann (LSZ) reduction formula then immediately follows
\begin{equation}
\lim_{\eta \to 0} \boldsymbol {\mathcal M}_k\big (\eta \vert \, \vec \mu(\vec q, \vec p) \big ) = {\mathcal M}_k\big (\vec \mu(\vec q, \vec p) \big ) \,,
\end{equation}
where the right-hand side here denotes the $2 \to n_k$ scattering amplitude, a function of $3 n_k - 4$ Lorentz invariants arranged in the vector $\vec \mu$.

We conclude by cancelling the common delta function in \Eq~\eqref{eq:AexpressionAPP} to reach
\begin{equation}
A_k(\vec q_1, \vec q_2; \vec p)= \delta_{k,1}
(2 \pi)^3 2 \omega_{\vec q_1} 2 \omega_{\vec q_2} \big [ \delta^3(\vec p_1 - \vec q_1) + \delta^3(\vec p_2 - \vec q_1) \big ] - 2 \omega_{\vec q_2} \frac{\boldsymbol {\mathcal M}_k\big (\eta(\vec q, \vec p) \vert \, \vec \mu(\vec q, \vec p) \big )^{\!*}}{ \eta( \vec q, \vec p) - i \epsilon}\bigg \vert_{q_1 + q_2 = P} \,,
\end{equation}
matching \Eq~\eqref{eq:Afinal} of the main text.

\subsection{General kinematics in the three-point function\label{app:kinematics}}

In this section we extend our expressions for the three-point function, presented in \Sec~\ref{sec:3pt}, to arbitrary kinematics. We begin by rewriting \Eq~\eqref{eq:CThetatoAB} as
\begin{equation}
\label{eq:cThetaCMframe}
c^{\Theta}_{\vec q_1 \vec q_2}(t_2 \vert \omega_0) = \sum_k \int \! d \Phi^\star_k \, e^{- ( E(\vec p) - \omega_{\vec q_1} - \omega_{\vec q_2}) t_2 } \, \Theta\big ( s(\vec p^\star)^{1/2} - 2 \omega_0, \Delta \big ) \, A^\star_k(\vec q_1^\star, \vec q_2^\star; \vec p^\star ) B^\star_k( \vec p^\star )
\,,
\end{equation}
where the $\star$ indicates that the quantity has been boosted with velocity $\vec \beta = - (\vec q_1 + \vec q_2)/(\omega_{\vec q_1} + \omega_{\vec q_2})$. To be more concrete we introduce the boost matrix $\Lambda(\vec \beta)$ which satisfies
\begin{equation}
{ \Lambda(\vec \beta)^\mu}_\nu \begin{pmatrix} \omega_{\vec q_1} + \omega_{\vec q_2} \\ \vec q_1 + \vec q_2 \end{pmatrix}^{\nu} = \begin{pmatrix} \sqrt{ \big ( \omega_{\vec q_1} + \omega_{\vec q_2} \big )^2 - \big ( \vec q_1 + \vec q_2 \big )^2 } \\ \vec 0 \end{pmatrix}^{\mu} = \begin{pmatrix} \omega_{\vec q_1}^\star + \omega_{\vec q_2}^\star \\ \vec 0 \end{pmatrix}^{\mu} \,.
\end{equation}
We then define
\begin{equation}
{ \Lambda(\vec \beta)^\mu}_\nu \begin{pmatrix} \omega_{\vec q_i} \\ \vec q_i \end{pmatrix}^{\nu} = \begin{pmatrix} \omega_{\vec q_i}^\star \\ \vec q_i^\star \end{pmatrix}^{\mu} \,, \qquad \qquad { \Lambda(\vec \beta)^\mu}_\nu \begin{pmatrix} \omega_{\vec p_i} \\ \vec p_i \end{pmatrix}^{\nu} = \begin{pmatrix} \omega_{\vec p_i}^\star \\ \vec p_i^\star \end{pmatrix}^{\mu} \,,
\end{equation}
which in turn allows us to introduce
\begin{align}
\begin{split}
A^\star_k(\vec q_1, \vec q_2; \vec p) & = \delta_{k,1} (2 \pi)^3 2 \omega^\star_{\vec q_1} 2 \omega^\star_{\vec q_2} \big [ \delta^3(\vec p^\star_1 - \vec q^\star_1) + \delta^3(\vec p^\star_2 - \vec q^\star_1) \big ] \\
& \hspace{160pt} - 2 \omega^\star_{\vec q_2} \frac{\boldsymbol {\mathcal M}_k\big (\eta(\vec q^\star, \vec p^\star) \vert \, \vec \mu(\vec q^\star, \vec p^\star) \big )^{\!*}}{ \eta( \vec q^\star, \vec p^\star) - i \epsilon} \,,
\label{eq:AfinalAppendix}
\end{split}\\
B^\star_k( \vec p^\star ) & = B_k( \vec p ) \,.
\end{align}
Then \Eq~\eqref{eq:cThetaCMframe} holds because $d \Phi_k 2 \omega_{\vec q_2}$, $(2 \omega_{\vec q_2})^{-1} A_k$ and $B_k$ are each Lorentz scalars.

It remains only to rewrite the $t_2$ dependence in the exponential. To do so we substitute
\begin{equation}
( E(\vec p) - \omega_{\vec q_1} - \omega_{\vec q_2}) t_2 = (P - q_1 - q_2)\cdot x_2 = ( E^\star(\vec p) - \omega^\star_{\vec q_1} - \omega^\star_{\vec q_2}) t^\star_2 \,,
\end{equation}
where we have introduced the four-vector $x_2^\mu = (t_2, \vec 0)$. The final line holds because $\vec q_1^\star + \vec q_2^\star = \vec 0$ and the Delta function inside $d \Phi_k$ then sets $\vec P^\star = \vec 0$.

Thus, we conclude that \Eqs~\eqref{eq:CThetatoABv2}, \eqref{eq:CThetaFinal}, \eqref{eq:CThetaThresh}, \eqref{eq:CThreshExp}, \eqref{eq:CThreshExpAlt}, \eqref{eq:CThetaGenEn}, \eqref{eq:finaCThetaThree}, \eqref{eq:CThetaAsymp} hold, also for $\vec q_1 + \vec q_2 \neq 0$ provided the following replacements are made
\begin{align}
\vec q & \longrightarrow \vec q_1^\star \,, \\
t_2 & \longrightarrow t_2^\star = \gamma t_2 \,,
\end{align}
where
\begin{equation}
\gamma = \frac{ \omega_{\vec q_1} + \omega_{\vec q_2}}{ \sqrt{ \big ( \omega_{\vec q_1} + \omega_{\vec q_2} \big )^2 - \big ( \vec q_1 + \vec q_2 \big )^2 } } \,.
\end{equation}

\subsection{Optical theorem for the three-point function\label{app:optical}}

In this section we demonstrate \Eq~\eqref{eq:Goptical} of the main text. Beginning with the definition of $\mathcal G(\mathcal E, \mathcal E/2)$
\begin{align}
\hspace{-20pt} \mathcal G(\mathcal E, \mathcal E/2) & = \sum_k \int d \Phi_k \, 2 \pi \delta \big( \mathcal E - E(\vec p) \big ) \, {\mathcal M}_k\big ( \vec \mu(\vec q, \vec p) \big )^{\!*} B_k(\vec p) \,, \\
& \hspace{-30pt} = \sum_k \frac{1}{S_k} \int \frac{d^3 \vec p_1}{(2 \pi)^3 2 \omega_{\vec p_1}} \cdots \frac{d^3 \vec p_{n_k}}{(2 \pi)^3 2 \omega_{\vec p_{n_k}}} \, (2 \pi)^4 \delta^3(\vec P ) \delta \big( \mathcal E - E(\vec p) \big ) \, {\mathcal M}_k\big ( \vec \mu(\vec q, \vec p) \big )^{\!*} B_k(\vec p) \,,
\label{eq:GdefApp}
\end{align}
we first apply the trick of \Reference~\cite{Maiani:1990ca} to write the scattering amplitude as a difference of inner-products
\begin{multline}
- i (2 \pi)^4 \delta^3(\vec P ) \delta \big( \mathcal E - E(\vec p) \big ) \, {\mathcal M}_k\big ( \vec \mu(\vec q, \vec p) \big )^{\!*} \\
= \langle \pi \pi, \text{in}, \vec q, -\vec q \vert k, \text{out}, \vec p_1 \cdots \vec p_{n_k} \rangle - \langle \pi \pi, \text{out}, \vec q, - \vec q \vert k, \text{out}, \vec p_1 \cdots \vec p_{n_k} \rangle \,.
\end{multline}
Inserting this into \Eq~\eqref{eq:GdefApp} then gives
\begin{align}
\label{eq:GStateSum}
\begin{split}
\mathcal G(\mathcal E, \mathcal E/2)
& = i \sum_k \frac{1}{S_k} \int \frac{d^3 \vec p_1}{(2 \pi)^3 2 \omega_{\vec p_1}} \cdots \frac{d^3 \vec p_{n_k}}{(2 \pi)^3 2 \omega_{\vec p_{n_k}}} \, \langle k, \text{out}, \vec p_1 \cdots \vec p_{n_k} \vert J(0) \vert 0 \rangle \\
& \hspace{10pt} \times \Big [ \langle \pi \pi, \text{in}, \vec q, -\vec q \vert k, \text{out}, \vec p_1 \cdots \vec p_{n_k} \rangle - \langle \pi \pi, \text{out}, \vec q, - \vec q \vert k, \text{out}, \vec p_1 \cdots \vec p_{n_k} \rangle \Big ] \,.
\end{split}
\end{align}
Finally we identify the combination of the integrals and the outer-product as an insertion of the identity on the Hilbert space
\begin{equation}
\mathbb{I} = \sum_k \frac{1}{S_k} \int \frac{d^3 \vec p_1}{(2 \pi)^3 2 \omega_{\vec p_1}} \cdots \frac{d^3 \vec p_{n_k}}{(2 \pi)^3 2 \omega_{\vec p_{n_k}}} \, \vert k, \text{out}, \vec p_1 \cdots \vec p_{n_k} \rangle \langle k, \text{out}, \vec p_1 \cdots \vec p_{n_k} \vert \,,
\end{equation}
such that \Eq~\eqref{eq:GStateSum} reduces to
\begin{align}
\mathcal G(\mathcal E, \mathcal E/2) & = i \Big [ \langle \pi \pi, \text{in}, \vec q, -\vec q \vert J(0) \vert 0 \rangle - \langle \pi \pi, \text{out}, \vec q, -\vec q \vert J(0) \vert 0 \rangle \Big ] \,, \\[8pt]
& =2 \, \text{Im} \, \langle \pi \pi, \text{out}, \vec q, -\vec q \vert J(0) \vert 0 \rangle \,,
\end{align}
thereby completing the proof.

\subsection{Cancellation of imaginary parts in the three-point function\label{app:cancel3}}

In this section we review the result of \Reference~\cite{Maiani:1990ca}, that the imaginary parts cancel between the first and second terms of \Eq~\eqref{eq:CThetatoABv2}, which we repeat here for convenience
\begin{align}
\label{eq:cTheta4ptRewrite}
c^{\Theta}_{\vec q, - \vec q}(t_2 \vert \omega_0) & = \Theta(2 \omega_q - 2 \omega_0, \Delta) \, f\big ( s( q^2) \big) + \Xi_{\vec q}(t_2 \vert \omega_0) \,, \\
\Xi_{\vec q}(t_2 \vert \omega_0)& \equiv - 2 \omega_q \int_{2 m_\pi}^\infty \! \frac{d \mathcal E}{2 \pi} \, e^{- ( \mathcal E - 2 \omega_{q}) t_2 } \,
\Theta( \mathcal E - 2 \omega_0, \Delta) \,
\frac{ \mathcal G(\mathcal E, \omega_q)}{ (\mathcal E - \omega_q)^2 - \omega_q^2 - i \epsilon} \,.
\end{align}

As a first step in the proof, we need to demonstrate that $\mathcal G(\mathcal E, \omega_q)$ is real, a result that we have shown in the previous section for $\mathcal G(\mathcal E, \mathcal E/2)$ and now extend for all values of the arguments. Beginning with the definition
\begin{equation}
\mathcal G(\mathcal E, \omega_q) = \sum_k \int d \Phi_k \, 2 \pi \delta \big( \mathcal E - E(\vec p) \big ) \, \boldsymbol{\mathcal M}_k\big (\eta(\vec q, \vec p) \vert \, \vec \mu(\vec q, \vec p) \big )^{\!*} B_k(\vec p) \,.
\end{equation}
We then substitute
\begin{multline}
- i (2 \pi)^4 \delta^3(\vec P ) \delta \big( \mathcal E - E(\vec p) \big ) \, \boldsymbol{\mathcal M}_k\big (\eta(\vec q, \vec p) \vert \, \vec \mu(\vec q, \vec p) \big )^{\!*} \\
= \langle \pi, \vec q; \tilde \pi(q), \text{in} \vert k, \text{out}, \vec p_1 \cdots \vec p_{n_k} \rangle - \langle \pi, \vec q; \tilde \pi(q), \text{out} \vert k, \text{out}, \vec p_1 \cdots \vec p_{n_k} \rangle \,,
\end{multline}
where we have introduced the shorthand
\begin{multline}
\langle \pi, \vec q; \tilde \pi(q), \text{in} \vert k, \text{out}, \vec p_1 \cdots \vec p_{n_k} \rangle \equiv \\
i \frac{[ \eta(\vec q, \vec p) - i \epsilon]}{\sqrt{Z_\pi}} \int d^4 x \, e^{i q^0 x^0 - i (- \vec q) \cdot \vec x} \langle \pi , \vec q \vert \pi(x) \vert k, \text{out}, \vec p_1 \cdots \vec p_{n_k} \rangle \,.
\end{multline}
This notation then allows us to directly imitate the derivation above to conclude
\begin{align}
\mathcal G(\mathcal E, \omega_q) & = i \Big [ \langle \pi, \vec q; \tilde \pi(-q), \text{in} \vert J(0) \vert 0 \rangle - \langle \pi, \vec q; \tilde \pi(- q), \text{out} \vert J(0) \vert 0 \rangle \Big ] \,, \\[8pt]
& =2 \, \text{Im} \, \langle \pi, \vec q; \tilde \pi(- q), \text{out} \vert J(0) \vert 0 \rangle \,,
\end{align}
from which directly follows $\mathcal G(\mathcal E, \omega_q) \in \mathbb R$.

Returning to \Eq~\eqref{eq:cTheta4ptRewrite}, we now evaluate the imaginary part of $ \Xi_{\vec q}(t_2 \vert \omega_0)$ to reach
\begin{align}
\text{Im} \, \Xi_{\vec q}(t_2 \vert \omega_0) & = - 2 \omega_q \text{Im} \int_{2 m_\pi}^\infty \! \frac{d \mathcal E}{2 \pi} \, e^{- ( \mathcal E - 2 \omega_{q}) t_2 } \, \Theta( \mathcal E - 2 \omega_0, \Delta) \,
\frac{ \mathcal G(\mathcal E, \omega_q)}{ (\mathcal E - \omega_q)^2 - \omega_q^2 - i \epsilon} \,, \\
& \hspace{-10pt} = - 2 \pi \omega_q \int_{2 m_\pi}^\infty \! \frac{d \mathcal E}{2 \pi} \, e^{- ( \mathcal E - 2 \omega_{q}) t_2 } \,
\Theta( \mathcal E - 2 \omega_0, \Delta) \, \mathcal G(\mathcal E, \omega_q) \delta \Big ( (\mathcal E - \omega_q)^2 - \omega_q^2 \Big ) \,, \\
& \hspace{-10pt} = - \Theta(2 \omega_q - 2 \omega_0, \Delta) \, \text{Im} f\big ( s( q^2) \big) \,.
\end{align}
Thus, the imaginary part of $\Xi$ cancels the imaginary part of $ \Theta(2 \omega_q - 2 \omega_0, \Delta) \, f\big ( s( q^2) \big)$, as claimed.

\subsection{Decomposition of the four-point function\label{app:details}}

In this section, we demonstrate \Eq~\eqref{eq:cThetaNpiv3} of the main text. We begin by recalling the basic definitions
\begin{multline}
\label{eq:CThetatoABNpiAPP}
c^{\Theta, N \pi}_{\vec q_1' \vec q_2' \vec q_1 \vec q_2}(t', t \vert \, M_0) \equiv \sum_k \int \! d \Phi_k \, e^{- ( E(\vec p) - \omega_{\vec q_1'} - \omega_{\vec q_2'}) t' } \, e^{ + ( E(\vec p) - \omega_{\vec q_1 } - \omega_{\vec q_2 }) t } \\[-5pt]
\times \Theta\big ( s(\vec p)^{1/2} - M_0, \Delta \big ) \, A_k(\vec q_1', \vec q_2'; \vec p ) A^*_k(\vec q_1, \vec q_2 ; \vec p ) \,,
\end{multline}
where $\omega_{\vec q_1} = \sqrt{m_\pi^2 + \vec q_1^2}$, $ \omega_{\vec q_2} = \sqrt{m_N^2 + \vec q_2^2}$, and
\begin{align}
A_k(\vec q_1', \vec q_2'; \vec p ) & = \delta_{k,1}
(2 \pi)^3 2 \omega_{\vec q_1'} 2 \omega_{\vec q_2'} \delta^3(\vec p_1 - \vec q_1') - 2 \omega_{\vec q_2'} \frac{\boldsymbol{\mathcal M}_k\big (\eta(\vec q', \vec p) \vert \, \vec \mu(\vec q', \vec p) \big )^{\!*}}{ \eta( \vec q', \vec p) - i \epsilon} \,.
\label{eq:AfinalNpiAPP}
\end{align}
The product of $A_k$ and $A_k^*$ generates four terms, schematically represented by
\begin{equation}
\Big ( \delta + \frac{1}{\eta + i \epsilon} \Big ) \Big ( \delta + \frac{1}{\eta - i \epsilon} \Big ) = \delta^2 + \delta \frac{1}{\eta - i \epsilon} + \frac{1}{\eta + i \epsilon} \delta + \frac{1}{\eta + i \epsilon} \frac{1}{\eta - i \epsilon}\,.
\end{equation}
The doubly disconnected term is discarded and, for the single delta function term, the integral is removed so that one can send $\epsilon \to 0$ to recover
\begin{equation}
\Big ( \delta + \frac{1}{\eta + i \epsilon} \Big ) \Big ( \delta + \frac{1}{\eta - i \epsilon} \Big ) \ \longrightarrow \ \delta \frac{1}{\eta} + \frac{1}{\eta } \delta + \frac{1}{\eta + i \epsilon} \frac{1}{\eta - i \epsilon} \,.
\end{equation}
As the final term is inside the integral, the $i \epsilon$ must be treated carefully. The first step is to break each pole into real and imaginary parts
\begin{equation}
\delta \frac{1}{\eta} + \frac{1}{\eta } \delta + \frac{1}{\eta + i \epsilon} \frac{1}{\eta - i \epsilon} = \delta \frac{1}{\eta} + \frac{1}{\eta } \delta +\bigg [ \text{Re} \frac{1}{\eta + i \epsilon} + i \text{Im} \frac{1}{\eta + i \epsilon} \bigg ] \bigg [ \text{Re} \frac{1}{\eta - i \epsilon} + i \text{Im} \frac{1}{\eta - i \epsilon} \bigg ] \,.
\end{equation}
In the analysis below we will consider $q \neq q'$ and only take the limit $q \to q'$ on combinations of terms for which we know this is safe. Indeed, for many of the individual terms here the limit does not exist.

After multiplying out the product of binomials on the right-hand side, one has a total of six contributions to $c^{\Theta, N \pi}_{\vec q_1' \vec q_2' \vec q_1 \vec q_2}$. We find it most instructive to introduce some additional notation for this decomposition
\begin{equation}
c^{\Theta, N \pi}_{\vec q_1' \vec q_2' \vec q_1 \vec q_2}(t', t \vert \, M_0)_{\sf c} = \sum_{\vec n = \vec 1}^{\vec 6} c^{\Theta, N \pi, [\vec n]}_{\vec q_1' \vec q_2' \vec q_1 \vec q_2}(t', t \vert \, M_0)\,,
\end{equation}
where schematically the correspondence is given by
\begin{equation}
\bigg \{ [\vec 1] = \delta \frac{1}{\eta}, \ \
[\vec 2] =\frac{1}{\eta } \delta , \ \
[\vec 3] = \text{Im} \times \text{Re}, \ \
[\vec 4] = \text{Re} \times \text{Im}, \ \
[\vec 5] = \text{Im} \times \text{Im}, \ \
[\vec 6] = \text{Re} \times \text{Re}
\bigg \} \,.
\end{equation}

For example, $c^{\Theta, N \pi, [\vec 1]}$ is given explicitly by
\begin{equation}
\begin{split}
c^{\Theta, N \pi, [\vec 1]}_{\vec q_1' \vec q_2' \vec q_1 \vec q_2} & \equiv - \int \! \frac{d^3 \vec p_1 }{(2 \pi)^3 2 \omega_{\vec p_1} 2 \omega_{\vec p_2}} \, e^{- ( E(\vec p) - \omega_{\vec q_1'} - \omega_{\vec q_2'}) t' } e^{ + ( E(\vec p) - \omega_{\vec q_1 } - \omega_{\vec q_2 }) t } \\[-5pt]
& \hspace{0pt} \times \Theta\big ( s(\vec p)^{1/2} - M_0, \Delta \big ) \,
(2 \pi)^3 2 \omega_{\vec q_1'} 2 \omega_{\vec q_2'} \delta^3(\vec p_1 - \vec q_1') \, 2 \omega_{\vec q_2} \frac{\boldsymbol{\mathcal M}\big (\eta(\vec q, \vec p) \vert \, \vec \mu(\vec q, \vec p) \big )}{ \eta( \vec q, \vec p) + i \epsilon} \,,
\end{split}
\end{equation}
where the $\delta_{k,1}$ projects us into a single two-particle channel and we drop the $k=1$ subscript. Using the Dirac delta function to evaluate the remaining integrals yields
\begin{multline}
c^{\Theta, N \pi, [\vec 1]}_{\vec q_1' \vec q_2' \vec q_1 \vec q_2} = - e^{ + ( \omega_{\vec q_1'} + \omega_{\vec q'_2} - \omega_{\vec q_1 } - \omega_{\vec q_2 }) t }
\Theta\big ( s(\vec q')^{1/2} - M_0, \Delta \big ) 2 \omega_{\vec q_2} \frac{\boldsymbol{\mathcal M}\big (\eta(\vec q, \vec q') \vert \, \vec \mu(\vec q, \vec q') \big )}{ \eta( \vec q, \vec q')} \,,
\end{multline}
or, in the center-of-mass frame with $\vec q_1 = - \vec q_2 = \vec q$,
\begin{equation}
c^{\Theta, N \pi, [\vec 1]}_{\vec q', \vec q} = - e^{ ( \mathcal E(q') - \mathcal E(q) ) t } \,
\Theta\big ( \mathcal E(q') - M_0, \Delta \big ) \, 2 \omega_{N,q} \frac{\boldsymbol{\mathcal M}\big (\eta( q, q') \vert \, \vec \mu(\vec q, \vec q') \big )}{ \eta( q, q')} \,.
\end{equation}
This matches the second line of \Eq~\eqref{eq:cThetaNpiv3}, except for the fact that here we have the complex valued $\boldsymbol{\mathcal M}$ rather than only its real part. This is because $c^{\Theta, N \pi, [\vec 3]} $ generates the exact cancellation of the imaginary part, as we now show.

For general kinematics, $c^{\Theta, N \pi, [\vec 3]}$ is given by
\begin{align}
\begin{split}
c^{\Theta, N \pi, [\vec 3]}_{\vec q_1' \vec q_2' \vec q_1 \vec q_2} & \equiv i \pi \sum_k \int \! d \Phi_k \, e^{- ( E(\vec p) - \omega_{\vec q_1'} - \omega_{\vec q_2'}) t' } e^{ + ( E(\vec p) - \omega_{\vec q_1 } - \omega_{\vec q_2 }) t } \, \Theta\big ( s(\vec p)^{1/2} - M_0, \Delta \big ) \\[-5pt]
& \hspace{40pt} \times
\, 2 \omega_{ \vec q_2'} \delta \big( \eta(\vec q', \vec p) \big ) \mathcal M_k\big ( \vec \mu(\vec q', \vec p) \big ) 2 \omega_{ \vec q_2} \mathcal P \frac{\boldsymbol {\mathcal M}_k\big (\eta(\vec q, \vec p) \vert \vec \mu(\vec q, \vec p) \big )^{\!*}}{ \eta( \vec q, \vec p) } \,,
\end{split}
\end{align}
and, in the case of back-to-back momenta, this reduces to
\begin{align}
\begin{split}
c^{\Theta, N \pi, [\vec 3]}_{\vec q', \vec q} & = i \pi \sum_k \int \! d \Phi_k \, e^{- (E(\vec p) - \mathcal E(q')) t' } e^{ + ( E(\vec p) - \mathcal E(q) ) t } \, \Theta\big ( \mathcal E(q') - M_0, \Delta \big ) \\[-10pt]
& \hspace{60pt} \times
\, \delta \big( E (\vec p)- \mathcal E(q') \big ) \mathcal M_k\big ( \vec \mu(\vec q', \vec p) \big ) 2 \omega_{N, q} \mathcal P \frac{\boldsymbol {\mathcal M}_k\big (\eta( q, p) \vert \vec \mu(\vec q, \vec p) \big )^{\!*}}{ \eta( q, p) } \,,
\end{split} \\[5pt]
& = \frac{ i 2 \omega_{N, q} \, \Theta\big ( \mathcal E(q') - M_0, \Delta \big )\, e^{ + ( \mathcal E(q') - \mathcal E(q) ) t } }{\eta(q,q')} \, \Xi(\vec q, \vec q') \,,
\end{align}
where we have used $\eta(\vec q', \vec p) = (E(\vec p) - \omega_{\pi, q'}) - \omega_{N, q'}^2 = [ E(\vec p) - \omega_{\pi, q'} + \omega_{N, q'} ] [E(\vec p) - \mathcal E(q')]$ for the case of back-to-back momenta. In addition we have defined
\begin{equation}
\Xi(\vec q, \vec q') \equiv \pi \sum_k \int \! d \Phi_k \, \delta (E(\vec p) - 2 \omega_{q'} ) \mathcal M_k\big ( \vec \mu(\vec q', \vec p) \big ) \boldsymbol{\mathcal M}_k \big (\eta( q, p) \vert \vec \mu(\vec q, \vec p) \big )^{\!*}\,.
\end{equation}

Next, following the original argument of \Reference~\cite{Maiani:1990ca}, reviewed in \App~\ref{app:optical}, one can show that $\Xi$ is just the imaginary part of $\boldsymbol {\mathcal M}$:
\begin{align}
\begin{split}
\Xi(\vec q, \vec q') & = - i \pi \sum_k \int \! d \Phi_k \delta (2 \omega_{q'} - E(\vec p)) \, \eta( q, p) \, 2 \omega_{\vec q_2} \frac{\langle \pi , \vec q_1 \vert \pi(0) \vert k, \text{out}, \vec p_1 \cdots \vec p_{n_k} \rangle}{\sqrt{Z_\pi}} \\
& \hspace{20pt} \times \Big [ \langle k, \text{out}, \vec p_1 \cdots \vec p_{n_k} \vert N \pi, \text{in}, \vec q, -\vec q \rangle - \langle k, \text{out}, \vec p_1 \cdots \vec p_{n_k} \vert N \pi, \text{out}, \vec q, - \vec q \rangle \Big ] \,,
\end{split} \\[8pt]
& = \frac{1}{2i} \frac{\eta( q, p) 2 \omega_{\vec q_2} }{\sqrt{Z_\pi}} \Big [ \langle \pi , \vec q_1 \vert \pi(0) \vert N \pi, \text{in}, \vec q, -\vec q \rangle - \langle \pi , \vec q_1 \vert \pi(0) \vert N \pi, \text{out}, \vec q, -\vec q \rangle \Big ] \,, \\[8pt]
& = \text{Im} \, \boldsymbol {\mathcal M}\big (\eta( q, q') \vert \, \vec \mu(\vec q, \vec q') \big ) \,.
\end{align}
Note that strictly one should substitute only the connected part of the matrix element for $\vec {\mathcal M}$. As the disconnected part is pure real, however, the final line holds as written. Setting $\eta( q, q') = 0$ on both sides of this equation also yields the result
\begin{equation}
h_0 =  \text{Im} \, \mathcal M\big ( \vec \mu(\vec q, \vec q') \big ) \,,
\end{equation}
used to reach \Eq~\eqref{eq:4ptGenKinFin} of the main text.

Putting everything together yields
\begin{align}
c^{\Theta, N \pi, [\vec 1]}_{\vec q', \vec q} +c^{\Theta, N \pi, [\vec 3]}_{\vec q', \vec q} & = - e^{- ( \mathcal E(q) - \mathcal E(q') ) t } \,
\Theta\big ( \mathcal E(q') - M_0, \Delta \big ) \, 2 \omega_{ N, q} \frac{\text{Re} \, \boldsymbol { \mathcal M} \big (\eta( q, q') \vert \, \vec \mu(\vec q, \vec q') \big ) }{ \eta(q,q') } \,.
\end{align}
In addition, an essentially identical analysis gives the result for $[\vec 2]$ and $[\vec 4]$
\begin{align}
c^{\Theta, N \pi, [\vec 2]}_{\vec q', \vec q} +c^{\Theta, N \pi, [\vec 4]}_{\vec q', \vec q} & = - e^{+(\mathcal E(q') - \mathcal E(q) ) t' } \,
\Theta\big ( \mathcal E(q) - M_0, \Delta \big ) \, 2 \omega_{ N, q' } \frac{\text{Re} \, \boldsymbol { \mathcal M} \big (\eta( q', q) \vert \, \vec \mu(\vec q', \vec q) \big ) }{ \eta(q',q) } \,,
\end{align}
where the only distinction is the exchange of $q \leftrightarrow q'$ and $t \leftrightarrow -t'$. To conclude the tracking of terms we note that
\begin{equation}
c^{\Theta, N \pi, [\vec 6]}_{\vec q', \vec q} = \mathcal C(t', t \vert \vec q', \vec q, \omega_0, \Delta) \,,
\end{equation}
as defined in \Eq~\eqref{eq:calC4ptdef} of the main text, and in addition that $ c^{\Theta, N \pi, [\vec 5]}_{\vec q', \vec q} = 0$. This final claim holds because each term is defined away from $q = q'$, with the $q \to q'$ limit to be performed only on the sum. In this prescription, the double Delta function defining $c^{\Theta, N \pi, [\vec 5]}_{\vec q', \vec q} $ cannot be satisfied. This completes the demonstration of \Eq~\eqref{eq:cThetaNpiv3}.

\bibliographystyle{JHEP}
\bibliography{refs.bib}

\end{document}